\documentclass[11pt]{article}   	
\usepackage{geometry} 
\usepackage{amsmath,slashed} 								
\usepackage{graphicx}	
\usepackage{amssymb}
\usepackage{cite}
\usepackage{color,xspace,listings}
\usepackage[hyperfootnotes=false]{hyperref}
\numberwithin{equation}{section}
\DeclareSymbolFont{extraup}{U}{zavm}{m}{n}
\DeclareMathSymbol{\vardiamond}{\mathalpha}{extraup}{87}
\usepackage{float}
\restylefloat{table}
\allowdisplaybreaks

\def\twomat[#1,#2][#3,#4]{\left( \begin{array}{cc} #1 & #2 \\ #3 & #4 \end{array} \right)}
\def\thv[#1,#2,#3]{\left( \begin{array}{c} #1 \\ #2 \\ #3 \end{array} \right)}
\def\twv[#1,#2]{\left( \begin{array}{c} #1 \\ #2 \end{array} \right)}

\def\ov{\overline}

\def\GeV{\ensuremath{\mathrm{GeV}}}

\newcommand{\SARAH}{{\tt SARAH}\xspace}

\newcommand{\MG}{{\tt MadGraph}\xspace}

\newcommand{\pythia}{{\tt pythia}\xspace}
\newcommand{\Delphes}{{\tt Delphes}\xspace}
\newcommand{\fastjet}{{\tt fastjet}\xspace}

\newcommand{\YODA}{{\tt YODA}\xspace}

\newcommand{\CHECKMATE}{{\tt CheckMATE}\xspace}
\newcommand{\MA}{{\tt MadAnalysis}\xspace}
\newcommand{\GAMBIT}{{\tt GAMBIT}\xspace}

\newcommand{\smodels}{{\tt SModelS}\xspace}

\newcommand{\python}{{\tt python}\xspace}

\newcommand{\hepmc}{{\tt hepmc}\xspace}
\newcommand{\hepmcV}{{\tt hepmc2}\xspace}

\newcommand{\HA}{{\tt HackAnalysis}\xspace}
\newcommand{\HEPData}{{\tt HEPData}\xspace}

\def\pt{\ensuremath{p_T}\xspace}
\def\ptmiss{\ensuremath{p_T^{\rm miss}}\xspace}
\def\ptmissv{\ensuremath{{\bf p}_T^{\rm miss}}\xspace}
\def\ptmissmu{\ensuremath{p_T^{\rm miss, \slashed{\mu}}}\xspace}
\def\invfb{\ensuremath{{\rm fb}{}^{-1}}\xspace}
\def\GeV{\ensuremath{{\rm GeV}}\xspace}

\title{Blank Title}

\date{}

\begin{document}

\begin{flushright}
\end{flushright}
\begin{center}

\vspace{1cm}
{\LARGE{\bf Long Dead Winos}}

\vspace{1cm}

\large{Mark D. Goodsell$^{a,1}$ \let\thefootnote\relax\footnote{${}^1$goodsell@lpthe.jussieu.fr}and
Lakshmi Priya${}^{b,2}$\footnote{${}^2$lakshmipriya2609@gmail.com}
 \\[5mm]}

{ ${}^a$\sl Laboratoire de Physique Th\'eorique et Hautes Energies (LPTHE),\\ UMR 7589,
  Sorbonne Universit\'e et CNRS, 4 place Jussieu, 75252 Paris Cedex 05, France. \\
 ${}^b$ Indian Institute of Science Education and Research Thiruvananthapuram (IISER TVM), \\
Maruthamala PO, Vithura, Thiruvananthapuram - 695 551, Kerala, India.}

\end{center}
\vspace{0.7cm}

\abstract{We describe a new code and approach using particle-level information to recast the recent CMS disappearing track searches including all run 2 data. Notably, the simulation relies on knowledge of the detector geometry, and we also include the simulation of pileup events directly rather than as an efficiency function. We validate it against provided acceptances and cutflows, and use it in combination with heavy stable charged particle searches to place limits on winos with any proper decay length above a centimetre. We also provide limits for a simple model of a charged scalar that is only produced in pairs, that decays to electrons plus an invisible fermion.}

\newpage
\setcounter{footnote}{0}

%------------------------------------------------------------------------------------------------------------							
\section{Introduction}
\label{introduction}

Classic searches for new physics at the Large Hadron Collider (LHC) involve the assumption of new heavy particles that are either stable or decay rapidly, either to a stable (hidden) particle or Standard Model (SM) states. With the LHC entering a precision phase for existing searches, there is now significant interest in looking for alternatives to this paradigm, where new discoveries could be hiding in plain sight. One such alternative involves long-lived particles (LLPs) that may decay inside the detectors. This has attracted significant interest, including a series of workshops \url{https://longlivedparticles.web.cern.ch/} and a community white paper \cite{Alimena:2019zri}. Several searches for such particles have already been undertaken using run 2 data, and are often highly complementary to prompt searches for the same models.

A heavy \emph{charged} particle can have a long lifetime if it has very weak couplings and/or little phase space in its decays. In the latter case, if it decays to a slightly lighter neutral particle, most of its energy will be carried away and it will seem to disappear inside a particle detector. Such scenarios are very common in models of physics Beyond the Standard Model, especially among heavy $SU(2)$ multiplets which have a neutral component; these would only be split in mass by electroweak effects/loops which are typically on the order $\mathcal{O} (100)$ MeV, meaning that the decay of the charged state is typically into the neutral one and a single pion. Such a particle would make an excellent dark matter particle, and indeed Minimal Dark Matter \cite{Cirelli:2005uq,Cirelli:2014dsa} falls into this category; in the case of an $SU(2)$ multiplet it resembles a wino of (split) supersymmetry \cite{Chen:1996ap,ArkaniHamed:2004fb,Ibe:2012hu,Hall:2012zp, Arvanitaki:2012ps, Citron:2012fg}. 
Such signatures are also ubiquitous in non-minimal scenarios \cite{Garny:2017rxs,Wang:2017sxx,Bharucha:2018pfu,Biswas:2018ybc,Belyaev:2018xpf,Borah:2018smz,Belanger:2018sti,Filimonova:2018qdc,Das:2020uer} and in a recent example were found to appear naturally within the context of dark matter models arising from Dirac gauginos \cite{Goodsell:2020lpx}, where limits from disappearing tracks were also discussed.

ATLAS \cite{Aaboud:2017mpt} published a search for this signature based on $36.1$ \invfb, with substantial recasting material including efficiencies (i.e. information about the proportion of events that are selected by the experiments due to the response of the detector rather than just the cuts applied); this material was used and then applied to other models \cite{Belyaev:2020wok} in a code available on the {\tt LLPrecasting} github repository \url{https://github.com/llprecasting/recastingCodes}. Very recently a conference note \cite{ATLAS:2021ttq} with the full run 2 dataset of $136$ $\invfb$ appeared.

On the other hand, CMS published two disappearing track searches: \cite{Sirunyan:2018ldc} with 38.4 \invfb of data from 2015 and 2016 and \cite{Sirunyan:2020pjd} with 101 \invfb from 2017 and 2018. These together provide the most powerful exclusion for disappearing tracks (which should be equivalent to the newly released \cite{ATLAS:2021ttq}). However, while there was substantial validation material, including cutflows and acceptances for events with one and two charged tracks separately, unlike for \cite{Aaboud:2017mpt} no efficiencies were provided. While the provided acceptances can be used in a simplified models approach for fermionic disappearing tracks, as in \cite{Goodsell:2020lpx} based on \smodels \cite{Kraml:2013mwa,Alguero:2020grj,Ambrogi:2017neo,Ambrogi:2018ujg,Khosa:2020zar} (which includes LLPs), it is a unique challenge to apply the results of this search to other models: the development of a strategy and code which can \emph{recast} the search is the subject of this paper.

Indeed the recasting of LHC searches so that they can be applied to models (or subsets of the parameter space) other than those originally considered by the experimental analysis is by now well developed. It is increasingly common for analyses to provide supplemental information for this purpose, often in the form of efficiencies or even pseudo-code; see \cite{Abdallah:2020pec} for a recent review and references therein. There now exist several popular frameworks for this purpose, which differ in their principal objectives: \GAMBIT \cite{Athron:2017ard,Athron:2018vxy,Kvellestad:2019vxm} is a scanning tool, designed to explore the likelihood space of a model, which can construct likelihoods for model points from collider data through its module {\tt ColliderBit} \cite{Balazs:2017moi}; \CHECKMATE \cite{Drees:2013wra,Kim:2015wza,Dercks:2016npn} aims to check models for exclusion against a wide variety of analyses, including now several LLP ones \cite{Desai:2021jsa}; \MA \cite{Conte:2012fm,Conte:2013mea,Conte:2014xya,Conte:2018vmg,Araz:2019otb,Araz:2020lnp} aims at providing a general analysis framework for examining data in detail, which can also be used to check models against many important analyses provided by users in its Public Analysis Database \cite{Dumont:2014tja,Fuks:2021wpe}. Notable also is {\tt rivet} \cite{Buckley:2019stt,Bierlich:2019rhm} which is mainly used to store Standard Model analyses for reuse, but can be applied to limit new physics scenarios (via their influence on SM processes) through {\tt contur} \cite{Butterworth:2016sqg,Buckley:2021neu}. 

There are also different strategies for detector simulation. Naively, the more complete the simulation the more accurate the recasting should be, but since the full simulation of ATLAS and CMS detectors remains closed-source the alternatives in practice are to use a fast simulation through \Delphes \cite{deFavereau:2013fsa}; or to use experimentally-determined efficiencies (which have been published for the LHC since before the beginning of operation, e.g.
\cite{Aad:2009wy})% or \url{https://cds.cern.ch/collection/CMS\%20Detector\%20Performance\%20Summaries})
, but must be updated as the detectors and algorithms are improved) for reconstruction of physics objects (electrons, muons, jets etc) known as a ``smearing'' or ``simplified fast simulation'' approach (see e.g. \cite{Buckley:2019stt,Araz:2020lnp} for recent discussions). The approach of \Delphes models the response of the calorimeters and propagation of particles in a magnetic field, while also making use of efficiencies. However for the analysis of interest here we require a modelling of the tracker system (in particular because the signal regions are defined \emph{in terms of the number of layers hit}), which is not available; and also a concept of the physical extent of the calorimeters/muon system, which is also absent at present in both approaches. Hence to implement our recasting we created a lightweight and fast code that uses particle-level information, that can easily be adapted to other analyses, that we call \HA. Ultimately we expect that the analysis described here will become available in the existing frameworks and that \HA should be useful for prototyping new features, hence the facetious name.

Note that whether the charged Standard Model particle is a pion or a lepton is not especially relevant for the disappearing track signature. For example, a scalar partner of the leptons in supersymmetric models, if it is nearly degenerate with the neutral fermionic partner of the gauge/higgs bosons (the neutralino) -- such as in a co-annihilation scenario for dark matter -- would also give a disappearing track. This case has not yet been considered in the experimental searches or  \cite{Belyaev:2020wok}, and so as an application of our results we shall investigate such a model here. It has the interesting peculiarity that only events with \emph{two} charged tracks occur, since they must be produced in pairs, in contrast to the wino/SU(2) multiplet case.

This paper is organised as follows. In section \ref{SEC:RECAST} we provide the details of how this search has been recast: the description of the cuts in sec.~\ref{SEC:cuts}; how we modelled the detector in sec.~\ref{SEC:detector}; how we modelled pileup in sec.~\ref{SEC:pileup}; and details about event simulation and how they affect the missing energy calculation in sec.~\ref{SEC:MET}. In section \ref{SEC:VALIDATION} we discuss the validation of our approach, with additional material in appendix \ref{APP:CUTFLOWS}. In sec.~\ref{SEC:WINOS} we reproduce the CMS exclusion for winos, and combine it with a recasting of heavy stable charged particle searches to place limits on winos up to infinite lifetime; $200$ GeV winos are excluded for all proper decay lengths above $2$cm, and $700$ GeV winos for all proper decay lengths above 20cm. In sec.~\ref{SEC:SCALARS} we apply our code to a model with charged scalars that are only produced in pairs and decay to an electron and a neutral fermion (equivalent to a right-handed slepton in supersymmetric models that is almost degenerate with a neutralino, such as in a co-annihilation region). Finally in sec.~\ref{SEC:HACKANALYSIS} we give some details about the recasting code \HA and how it can be used. 

\section{Recasting the CMS disappearing track search}
\label{SEC:RECAST}

The CMS search for disappearing tracks interpreted its results in terms of supersymmetric models, and provided parameter cards to generate events alongside substantial recasting material for such models. Hence to recast and validate the analysis we must consider the same model(s) and reproduce their data. The relevant particles in this model are a  ``chargino,'' that is a charged Dirac fermion $\tilde{\chi}^\pm;$ and a ``neutralino,'' i.e. a stable neutral fermion $\tilde{\chi}^0.$ They are always produced in pairs, either of two charged particles $\tilde{\chi}^\pm \tilde{\chi}^\mp$ or one charged and one neutral, $\tilde{\chi}^\pm \tilde{\chi}^0$ (or two neutral fermions that leave no tracks). The chargino then produces a track in the detector when it lives long enough, and then ``disappears'' by decaying to a neutralino and a pion, or a lepton/neutrino pair:
\begin{align}\label{EQ:DECAYMODES}
 \tilde{\chi}^\mp \rightarrow \tilde{\chi}^0 + \pi^\mp, \qquad \tilde{\chi}^\mp \rightarrow \tilde{\chi}^0 + \ell^\mp + \nu_\ell,
 \end{align}
 hence events are characterised by the number of charginos present as one or two track events. 
 The production is considered to be purely electroweak in origin (i.e. any additional particles that could decay to them should be so heavy as to give negligible contribution, unlike e.g. the ``strong production'' scenario of \cite{Aaboud:2017mpt}) which means that initial state radiation (of the incoming partons) is relevant but final state radiation much less so.

The two scenarios considered were a wino and a higgsino; in general in supersymmetry the electroweakino sector consists of winos, higgsinos and binos which mix. A pure wino is an $SU(2) $ triplet, while a higgsino is a doublet and a bino a singlet. In the case that they are pure states with no mixing the mass splitting of the neutral and charged states can be calculated accurately at two loops \cite{Ibe:2012sx} and the decay width of the chargino into a single pion becomes only a function of its mass. From the experimental point of view, however, the actual mass splitting is not measurable, and only the lifetime is relevant. In the two scenarios considered by CMS, the decay channel in eq.~(\ref{EQ:DECAYMODES}) was either 100\% into pions (in the wino case), or 95\% pions and 5\% leptons (in the higgsino case). These differences are negligible from the recasting point of view. The production cross-sections for the wino case and higgsino case will be different functions of the mass, and the ratio of single track to double track events are different, roughly $2:1$ for winos and $7:2$ for higgsinos. However, since the recasting acceptances are given separately for double and single tracks, the wino and higgsino data are effectively identical. Hence throughout we shall focus only on the wino case.

While a pure wino or pure higgsino would have a lifetime given only by its mass, because the lifetime depends very strongly on the mass splitting it is very sensitive to a small admixture of other multiplets. Since a pure electroweak multiplet in supersymmetric models is only ever an approximation, and typically the mixing among the charginos or neutralinos is not negligible, in general the lifetime of the chargino can vary over many orders of magnitude, often without noticeably affecting the production cross-section or any of the other data relevant for the recasting. Hence to derive more general limits CMS were justified in varying the proper decay length of the chargino in each parameter point between $0.2$cm and $10000$cm by hand, and we shall therefore use the same parameter cards/approach here. Of course, more mixing between the states would break this assumption and so to test a fully general model it would be necessary to use a code such as the one described here. 

In this section we shall first describe the details of the experimental cuts, before presenting our approach to modelling the detector, pileup and finally validation of our code.

\subsection{Triggers and cuts}
\label{SEC:cuts}

The analyses \cite{Sirunyan:2018ldc,Sirunyan:2020pjd} require a hardware trigger on missing transverse energy (MET). However, one of the peculiarities of this type of search is that long-lived charged particles that reach the muon system will be reconstructed as muons, and therefore they lead to no missing energy. To mitigate this, \ptmissmu is used in the triggers and event selection, which is the missing momentum without including muons in the calculation. The exact triggers used in the analysis are not given, and it is known that the trigger thresholds actually changed throughout the data taking periods, which leads to differences in the provided cutflows between the 2017 and first part of 2018 data (that should naively be identical).

\begin{table}\centering
  \begin{tabular}{|c|c|} \hline
   Trigger & Period  \\ \hline
    $\ptmiss > 90\ \GeV$ or $\ptmissmu > 90\ \GeV$ & 2015 \\
    $\ptmiss > 120\ \GeV$ or $\ptmissmu > 120\ \GeV$& 2016, 2017,2018 \\
    $\ptmiss > 75\ \GeV$, isolated track with $\pt > 50\ \GeV$ & 2015, 2016 \\
    $\ptmiss > 105\ \GeV$, isolated track with $\pt > 50\ \GeV$ with $> 5$ hits & 2017, 2018\\ \hline
  \end{tabular}
  \caption{\label{TAB:Triggers} Triggers listed in the analysis papers, and therefore implemented in the code, along with the period to which they apply.}
\end{table}

The MET cuts (after the triggers) are then $\ptmiss > 100\ \GeV$ for the 2015/16 data and $\ptmissmu > 120\ \GeV$ for 2017/18. To match the first few cuts in the provided cutflows it is necessary in reconstruction to take into account the probability of the triggers turning on, for example from \cite{Sirunyan:2019kia}. In the simulation we use an approximation based on the information there.

Subsequently, there are cuts based on the jets and missing energy:
\begin{itemize}
\item We require at least one jet with $p_T > 110$ GeV and $|\eta| < 2.4$. Jets are reconstructed using the anti-$k_T$ algorithm \cite{Cacciari:2008gp,Cacciari:2011ma} with radius parameter $R=0.4$.
\item The difference $\Delta \phi$ between the highest-$p_T$ jet and \ptmissv must be greater than $0.5$ radians.
\item If $n_{jets} \ge 2$, the maximum $\Delta \phi_{\rm max} < 2.5$ radians.
\item Due to a failure in 2018, we reject $-1.6 < \phi(\ptmissv) < -0.6$ for the second part of the data taking period in 2018B (39 \invfb) known as the ``HEM veto.''
\end{itemize}

After the cuts, the selection of charged tracks are weeded out:
\begin{itemize}
\item We keep only isolated tracks with $\pt > 55\ \GeV$ and $|\eta|<2.1$; isolation is such that the scalar sum of the \pt of all other tracks within $\Delta R < 0.3$ is less than 5\% of candidate track's \pt.
\item We remove tracks within regions of incomplete detector coverage in the muon system for $0.15 < | \eta | < 0.35$ and $1.55 < | \eta | < 1.85$, and $1.42 < |\eta | < 1.65 $ corresponding to the transition region between the barrel and endcap sections of the ECAL.
\item Tracks whose projected entrance into the calorimeter is within $\Delta R < 0.05$ of a nonfunctional or noisy channel are rejected. The location of these are not given, nor is the total angular coverage given in the papers; it does not seem to be significant, however, and we shall ignore it. 
\item Tracks must be separated from jets having $\pt > 30 \GeV$ by $\Delta R ( {\rm track,\ jet} ) > 0.5$.
\item With respect to the primary vertex, candidate tracks must have a transverse impact parameter ($|d_0|$)
  less than 0.02 cm and a longitudinal impact parameter ($|d_z|$) less than 0.50 cm.
\item Tracks are rejected if they are within $\Delta R( {\rm track,\ lepton} ) < 0.15$ of any reconstructed lepton candidate, whether electron, muon, or $\tau_h$. This requirement is referred to as the “reconstructed lepton veto”. In particular, the overlap removal of LLPs near muons will apply to an LLP that is reconstructed as a muon.
\item Candidate tracks are rejected from the search  region if they are within an $\eta-\phi$ region in which the local efficiency is less than the overall mean efficiency by at least two standard deviations; this removes 4\% of remaining tracks, but there is no information about where they are -- we therefore may simply apply a flat $96\%$ probability of meeting this criterion for each track.
\end{itemize}

Next, the data are split into regions (2015, 2016A and 2016B, 2017, 2018A and 2018B) according to the cuts above (notably the different triggers/MET cut between the two analyses) and the HEM veto for 2018B, but also
\begin{itemize}
\item For the 2017 data, tracks are rejected within the angular region $ 2.7 < \varphi < \pi$, $0 < \eta < 1.42.$
\item For the 2018 data period, tracks are rejected within the angular region $ 0.4< \varphi < 0.8$, $0 < \eta < 1.42.$
\end{itemize}
We therefore see that the 2017 and 2018A periods have almost identical cuts (since the cut $\Delta \varphi \sim 0.4$ in both cases and the HEM veto applies to 2018B). There are differences in the experimentally provided cutflows which must instead be due to differences in triggers and trigger efficiencies; but the final efficiencies passing all cuts are very similar. There is also no information about the difference between periods 2016A and 2016B other than that the trigger configurations changed between the two in an unspecified way (which is unlikely to significantly affect the final efficiencies), so we treat them as identical in the analysis.

For each period, there are hit-based quality requirements on the tracks, which are detected as they pass through pixel and then tracker layers; and we must determine whether they have disappeared:
\begin{itemize}
\item Tracks must hit all of the pixel layers (three for 2015/16, four for 2017/18).
\item Tracks must have at least three missing \emph{outer} hits in the tracker layers along its trajectory, i.e. it must stop, and not just disappear just before leaving the tracker.
\item Track must have no missing \emph{middle} hits, i.e. there must be a continuous line of hits until it disappears.
\item $E_{\rm calo}^{\Delta R < 0.5} < 10\ \GeV$ for each track, i.e. the calorimeter energy measured within $\Delta R < 0.5$ must be less than $10\ \GeV$, ensuring that it does indeed stop (and is not just missed by the tracker). This cut is of negligible importance to signal events (but very important for background, which includes charged hadrons); it is also somewhat tricky to implement when considering pileup, since the pileup contribution to this measure must be subtracted, as will be discussed later. 
\end{itemize}

Finally, we must decide which signal region the tracks fall into; these are based on the number of layers hit by the track. For the 2017/2018 data, the three regions are:
\begin{itemize}
\item SR1: $n_{\rm lay} = 4$
\item SR2: $n_{\rm lay} = 5$
\item SR3: $n_{\rm lay} \ge 6$.  
\end{itemize}
For the original analysis/data in 2015 and 2016, there is only one signal region, corresponding to seven hits or more overall in the tracker. Since the earlier analysis took place before the pixel detector upgrade, the pixel detector contained only three layers whereas for the second analysis we have four. For the sake of simplicity we only model the latter detector, and therefore we impose the criterion of more than seven hits for these regions.

\subsection{Modelling the detector}
\label{SEC:detector}

Recent experimental analyses often include techniques for compensating for the imperfections of the detectors, to the point that simulating the detector response as performed e.g. in {\tt Delphes} may be less accurate than using particle-level information and data-driven efficiencies. For this analysis, the most important physics objects are the charged tracks, muons, the MET and jets. Of these, the momenta/energies of the jets are relatively unimportant, in that we require only the directions; and the efficiency of the MET reconstruction has been studied \cite{Sirunyan:2019kia}. On the other hand, there is no public code to simulate the tracker response to charged particles in terms of hits and layers, and for LLP analyses, and this one in particular, we must take into account the physical dimensions of the detector: whether a charged particle decays within before leaving the muon system, for example, so that we can properly compute the MET. 

There is no efficiency information given by the experiment about the reconstruction of tracks as a function of length or direction. Moreover, the signal regions are defined in terms of the number of pixel/tracker layers traversed. This means that to accurately recast the analysis the code must have some model of the position of the layers. Moreover, by using a probability of a layer hit being registered we can model the possibility of missing inner/middle hits, as well as ``fake'' outer hits, and this appears to be important for accurately reproducing the signal efficiencies, which are otherwise too high; we use a fixed per-hit efficiency of $94.5\%$ \cite{Veszpremi:2017yvj} and clearly this has a significant impact on reducing the number of tracks observed, because we just need to miss one hit along the length of the track for it to be discounted.

Otherwise, to model the track hits, we assign a track object to each charged particle passing through the tracker and then compute the layers that it may pass through. We model the location of the layers in the pixel, inner barrel and outer barrel as cylinders, and inner discs and end-cap discs as fixed radius discs at given transverse distances from the interaction point, assumed to be perpendicular to the beam axis (which is not exactly correct, but good enough). The geometry and positions, along with a host of other useful information for this search, are found in \cite{Chatrchyan:2008aa,CMS:2012sda,Veszpremi:2017yvj} and especially the thesis of Adam Hart, \url{http://rave.ohiolink.edu/etdc/view?acc_num=osu1517587469347379} and references therein.

\begin{table}\centering
  \begin{tabular}{||c||ccccccccc||} \hline\hline
    Component & \multicolumn{9}{c||}{Layer positions (mm)} \\ \hline
    % TIBR= {230.0, 300.0, 400.0,500.0};
    Inner Barrel (radii) & 230 & 300 & 400 & 500 & &&& & \\
    % TIDZ= {775.0,900.0,1025.0};
    % TOBR = {608.0,692.0,780.0,868.0,965.0,1080.0};
    % TECZ= {1250.0,1400.0,1550.0,1700.0,1950.0,2000.0,2225.0,2450.0,2700.0};
    Inner discs (z) & 775 & 900 & 1025 &  & &&&& \\
    Outer barrel (radii) & 608 & 692 & 780 & 868 &965 &1080 & & & \\
    Endcaps (z) & 1250 &1400 & 1550 &1700 & 1950 & 2000 &2225 &2450 &2700 \\ \hline
  \end{tabular}
\caption{\label{TAB:LayerLocations} Locations used for the layers in the CMS tracker.}
\end{table}

\subsection{Effect of pileup and track isolation}
\label{SEC:pileup}

In principle, pileup affects the jets, MET calculation, track isolation and calculation of $E_{\rm calo}^{\Delta R < 0.5}$, as well as providing fake tracks.  To compensate, CMS employ rather effective mitigation techniques. Moreover, simulating pileup in a simplified simulation is rather unusual, since its effects are usually absorbed into the efficiencies. However, we have no such efficiencies that apply for this analysis, except for the MET trigger. Therefore we implemented pileup events in the code \HA. This is accomplished by simulating minimum bias events in \pythia, and storing the final state particles that register in the detector, and also metastable particles that leave tracks. These mainly consist of hadrons. These events are stored in a compressed text file, which can then be read into \HA (in this way we save significant processing time). As each signal event is read in/simulated, a number, drawn from a Poisson distribution with mean equal to the observed pileup average, of these stored events is randomly selected from the minimum bias database, and added to the event, with their vertices randomly distributed along the beam axis and in time according to the same distribution as used in \Delphes pileup events.

In general in a simplified simulation approach there are then a number of different ways for pileup events to be combined with the signal event. In a naive detector simulation we would combine them before any jet clustering and the MET calculation was performed. However, sophisticated techniques are employed by the experimental collaborations such as Jet Vertex Tagging to remove the effects of pileup, and we would then be forced to attempt to implement some version of these without knowledge of the details. Instead, we can choose to only include these events in the event record after the MET and jet clustering calculations have been performed. In this way they can be used to provide fake tracks and be included in the isolation computation. However, the idea of \HA is that the user can modify these aspects to suit their use case.

In addition, when particle-level data (as opposed to detector simulation) is used in codes such as GAMBIT, typically the effect of isolation requirements upon particle reconstruction are incorporated via efficiencies derived from experimental data. However, for our disappearing charged tracks we have no such efficiencies and so we store the momenta of the hadrons (both charged and neutral, that are long-lived enough to leave a track/reach the colorimeters) in the event record so that they can be used to compute the isolation. This is different to the typical approach of forgetting about hadrons once they have been clustered into jets. However, by examining the cutflows in appendix \ref{APP:CUTFLOWS} it is clear that the isolation requirement imposes a very severe reduction in the number of events (of around $60\%$), and so this inclusion is absolutely vital; it can also be seen that our modelling works rather well. 

For the CMS disappearing track analysis considered here, we chose to include the effects of pileup only for the the isolation calculation and for providing fake tracks (i.e. hadrons that are identified as disappearing tracks, potentially because of missing hits). Ideally we would also include its effects in the calculation of  $E_{\rm calo}^{\Delta R < 0.5},$ however even after subtracting the median energy (which is the standard mitigation technique) its effects were still too large and we simply disabled this cut. As can be seen from the cutflows in table \ref{TAB:PILEUP} where we compare the simulation with and without pileup events it is actually the track isolation that plays the biggest role in identifying the track as having disappeared, and the $E_{\rm calo}^{\Delta R < 0.5}$ cut has negligible effect. It is also clear from comparing the cutflows that the effects of pileup are tiny, so that in principle an almost as accurate result could be found by neglecting it in this case.

\begin{table}\centering
  \begin{tabular}{||c||c||c|c|c||c|c|c||} \hline \hline
     &  & \multicolumn{3}{|c||}{Expected Background}& \multicolumn{3}{|c||}{Observed Events} \\
    Period & Integrated Luminosity (\invfb) & $n_{\rm lay} =4 $ & $n_{\rm lay} =5 $  & $n_{\rm lay} \ge 6 $ & $n_{\rm lay} =4 $ & $n_{\rm lay} =5 $  & $n_{\rm lay} \ge 6 $ \\ \hline\hline
    2015 & $2.7$& -- & -- & $0.1 \pm 0.1$ & --& --& 1\\ \hline
    2016A & $8.3$ & --& --& $2.4 \pm 0.6$& --& --& 2\\ \hline
    2016B & $27.4$ & -- & -- & $4.0 \pm 1.1$&-- &-- &4\\ \hline
    2017 & $42.0$& $12.2 \pm 4.8$& $2.1 \pm 0.7$& $6.7 \pm 1.3$& 17& 4&6\\ \hline
    2018A & $21.0$&$7.3 \pm 3.5$ & $0.6 \pm 0.3$ & $1.8\pm 0.6$& 5& 0&2\\ \hline
    2018B & $39.0$ & $10.3 \pm 5.4$& $1.0 \pm 0.3$& $5.7\pm 1.3$& 11&2 &1\\ \hline \hline
  \end{tabular}
  \caption{\label{TAB:RegionData} Data-taking periods, integrated luminosity, background and observed events.}
\end{table}

\subsection{Event simulation and MET calculation}
\label{SEC:MET}

To simulate events, the CMS analysis used leading-order \pythia \cite{Sjostrand:2014zea} simulations. These cannot properly account for hard initial state radiation (ISR); therefore CMS adjusted the MET and the $\pt$ of the chargino pair using a data-driven approach, by comparing the process to $Z\rightarrow \mu\mu$ events. This is somewhat difficult to implement in a recasting tool, and in any case for different models it may not be applicable, since the topology of the process may be very different. Hence we take the (more standard) approach of simulating the hard process including up to two hard jets using \MG \cite{Alwall:2014hca}. However, there is then the choice of how to match to the parton showers, since there are several prescriptions. To investigate the effects of this, we implemented three different approaches: MLM matching \cite{Mangano:2001xp} within \pythia; CKKW-L merging \cite{Lonnblad:2001iq,Lonnblad:2011xx,Lonnblad:2012ix} within \pythia; and the version of MLM matching using reweighting from \MG 2.9 (which also uses \pythia). This latter approach meant using a \hepmc \cite{Dobbs:2001ck} interface, and was substantially slower to run per point, hence we used it for the benchmark points but not for a complete scan. 

\begin{figure}\centering
  \includegraphics[width=0.4\textwidth]{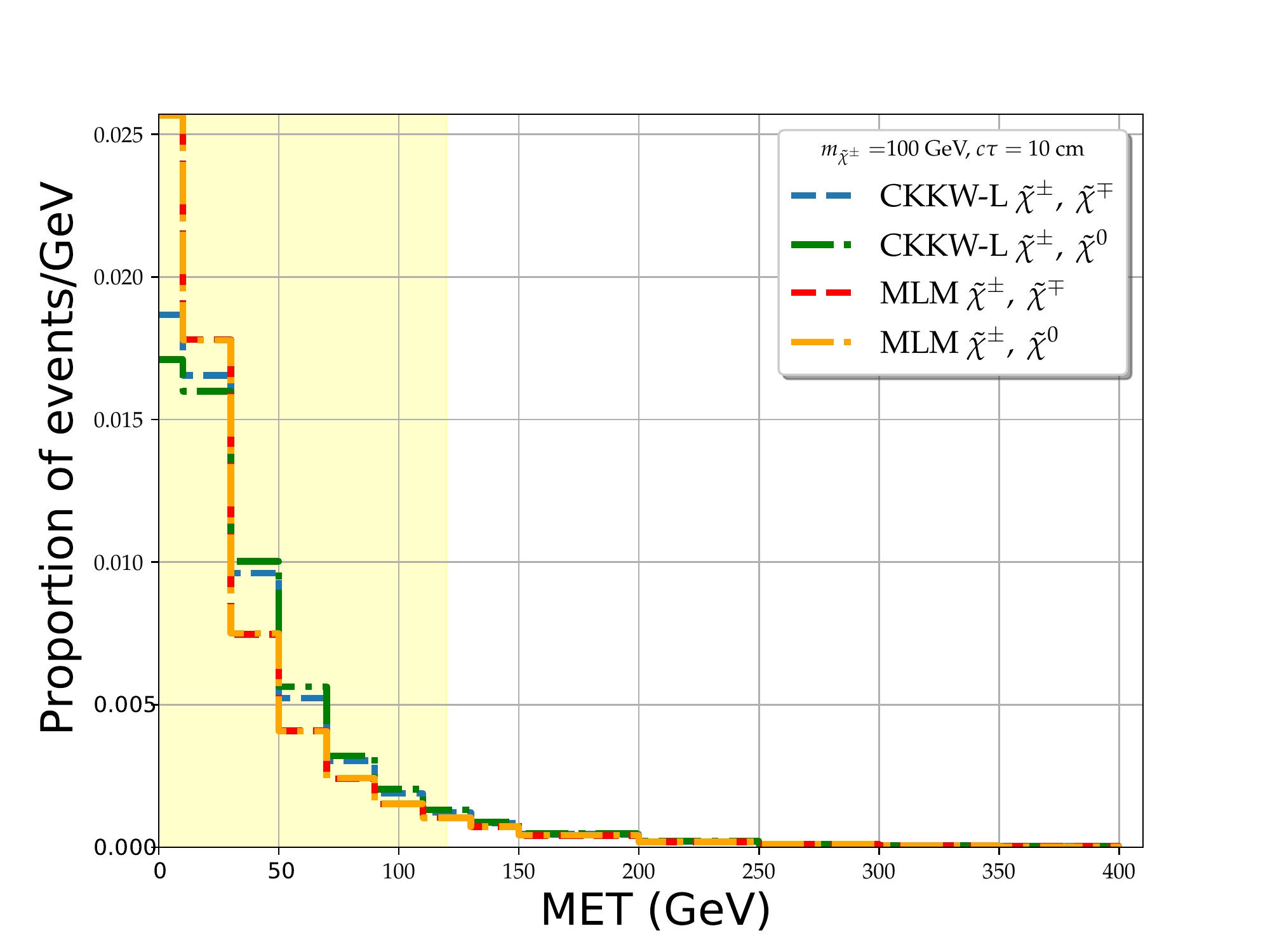} \includegraphics[width=0.4\textwidth]{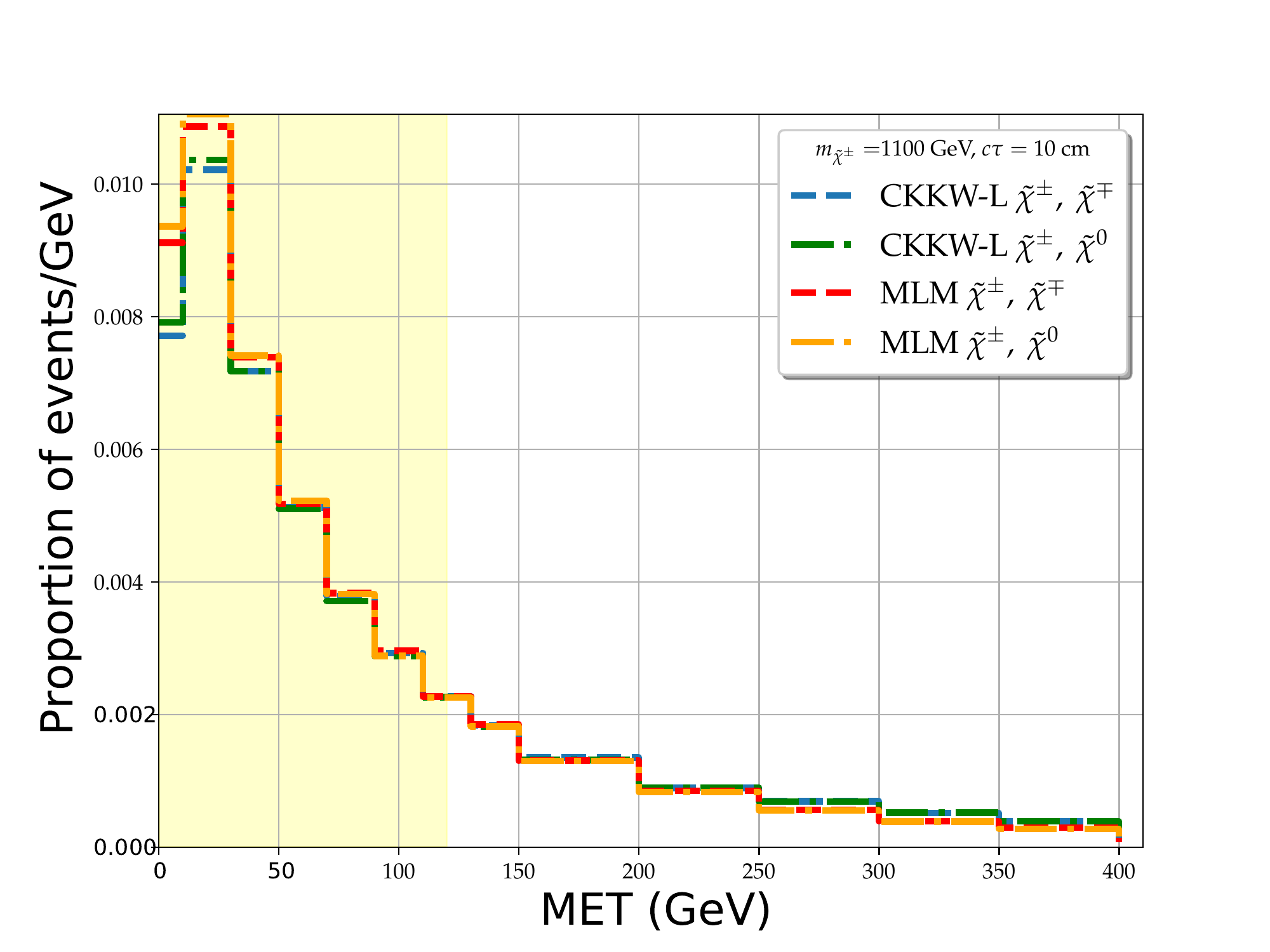}
  \caption{\label{FIG:METlowtau}Comparison of missing transverse energy distribution (proportion of events per GeV in each bin) for double-chargino and single-chargino events, for both CKKW-L merging and MLM matching. Left: $100$ GeV winos with 10cm proper decay length; right: $1100$ GeV winos with 10cm proper decay length. Not shown is the overflow bin.}
\end{figure}

The effect of the different matching/merging approaches should be seen in the MET calculation and also the distribution of momenta of the tracks. While they should be equivalent, CKKW-L merging is generally regarded as superior (if more complicated) because it leads to a smoother distribution; but provided that the matching scale in MLM is well chosen (and related to the hard process) this should not be a problem. We took the merging/matching scale to be one quarter of the chargino mass. In figure \ref{FIG:METlowtau} we compare the merging/matching approaches for $100$ GeV and $1100$ GeV. Clearly the distributions are both smooth, and there is not an obvious merit for one or the other from the plots; the differences are also small. However, since the MET cut and even more so the trigger appear on a rapidly falling part of the distribution, the small differences are amplified and the proportion of events passing the cut differ by between $10\%$ to $30\%$, which leads to substantial uncertainty on the predictions. 

\begin{figure}\centering
  \includegraphics[width=0.4\textwidth]{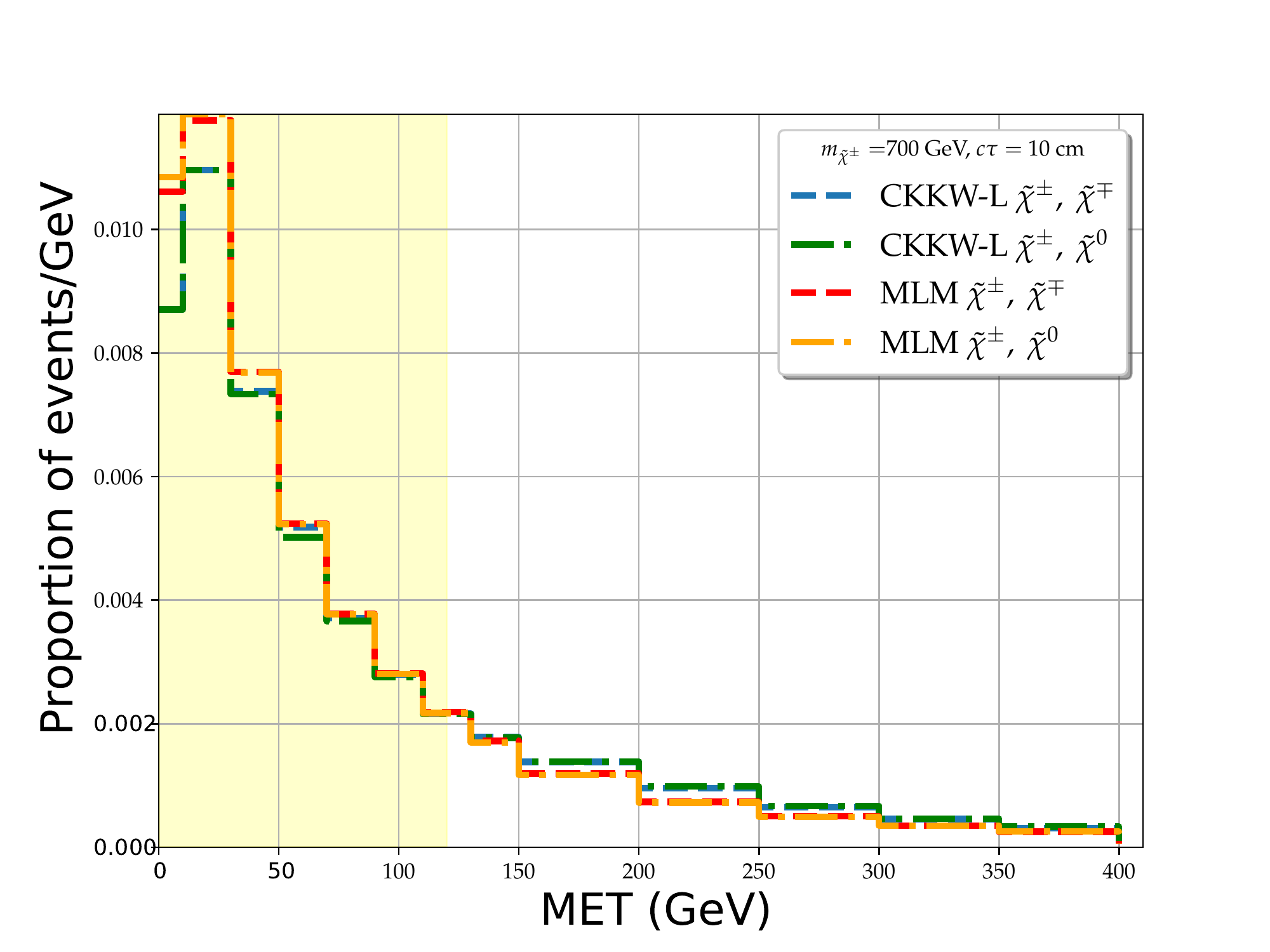} \includegraphics[width=0.4\textwidth]{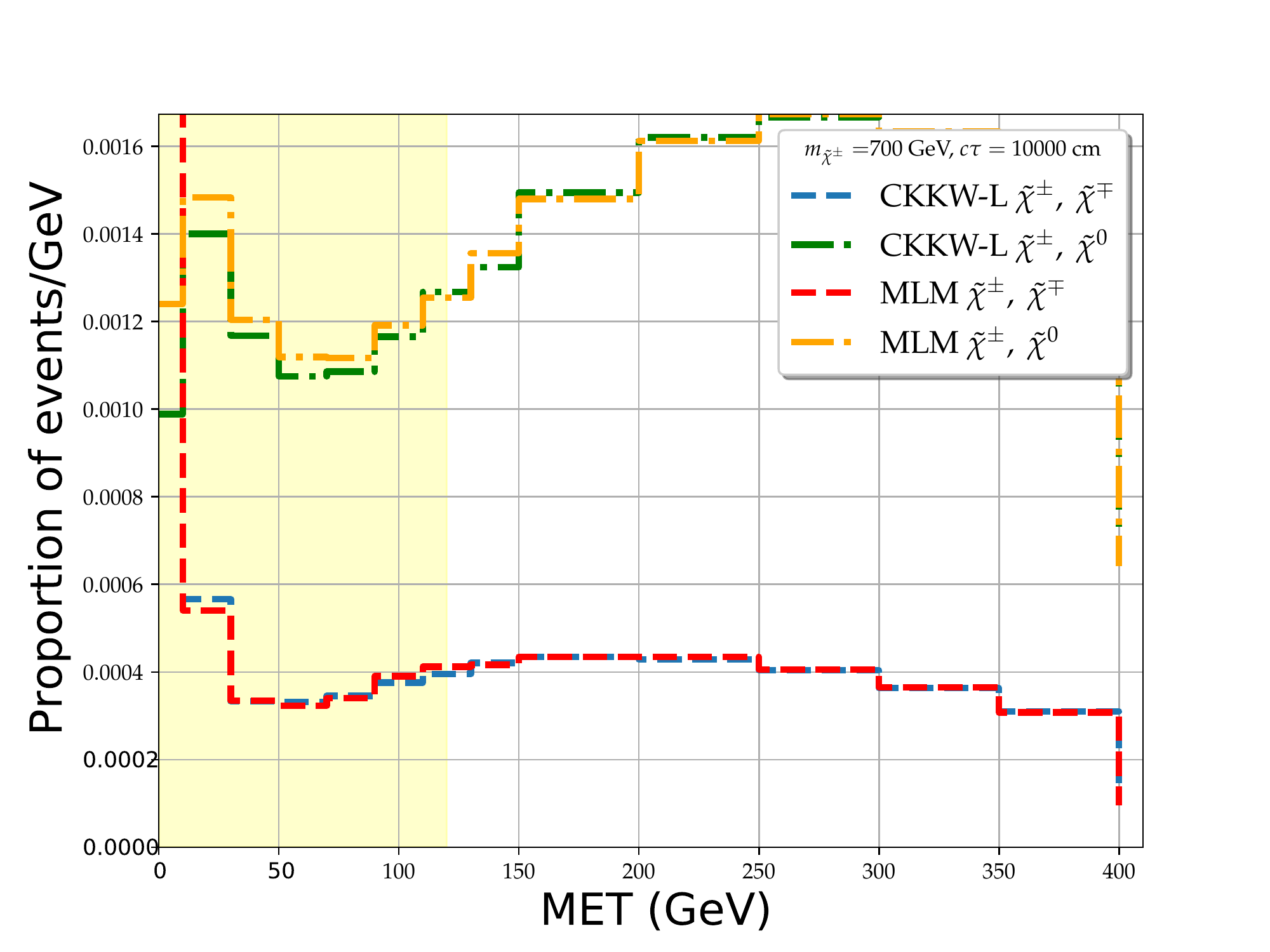}
  \caption{\label{FIG:MET700}Comparison of missing transverse energy distribution (proportion of events per GeV in each bin) for double-chargino and single-chargino events, for both CKKW-L merging and MLM matching. Left: $700$ GeV winos with 10cm proper decay length; right: $700$ GeV winos with 10000cm proper decay length. On the right-hand figure, nearly $80\%$ of $\tilde{\chi}^\pm \tilde{\chi}^\mp$ events lie in the \emph{first} bin, since the charginos are both escaping and being classed as muons, giving almost no missing energy. For $\tilde{\chi}^\pm \tilde{\chi}^0$ events the chargino nearly always escapes, leading to $40\%$ of the events having MET \emph{larger} than $400$ GeV; the overflow bin is not shown. }
\end{figure}

Not related to the impact of merging/matching, but to illustrate the effect of charginos escaping from the detector on the MET calculation, in figure \ref{FIG:MET700} the distribution of missing energy can be seen for 700 GeV winos with proper decay lengths of $10$cm and $10000$cm; in the right-hand plot most charginos continue to the muon system, so the single-chargino events have large missing energy, and double-chargino events even more so (i.e. most of the double-chargino events are in the overflow bin). This underlines the sensible choice of CMS to use both $\ptmiss$ and $\ptmissmu$ in the cuts, and that our code is able to reproduce this effect. 

\begin{figure}\centering
  \includegraphics[width=0.45\textwidth]{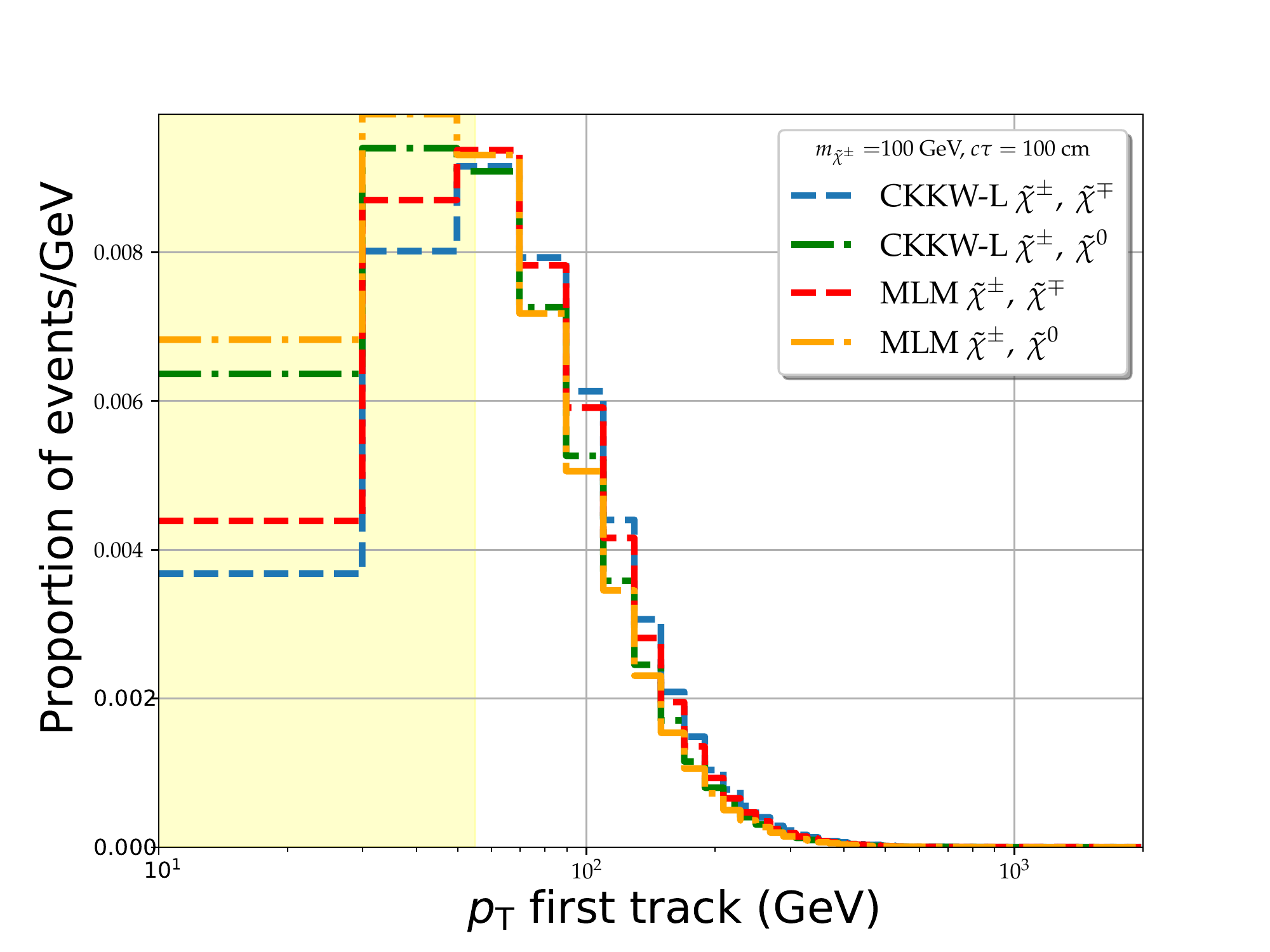} \includegraphics[width=0.45\textwidth]{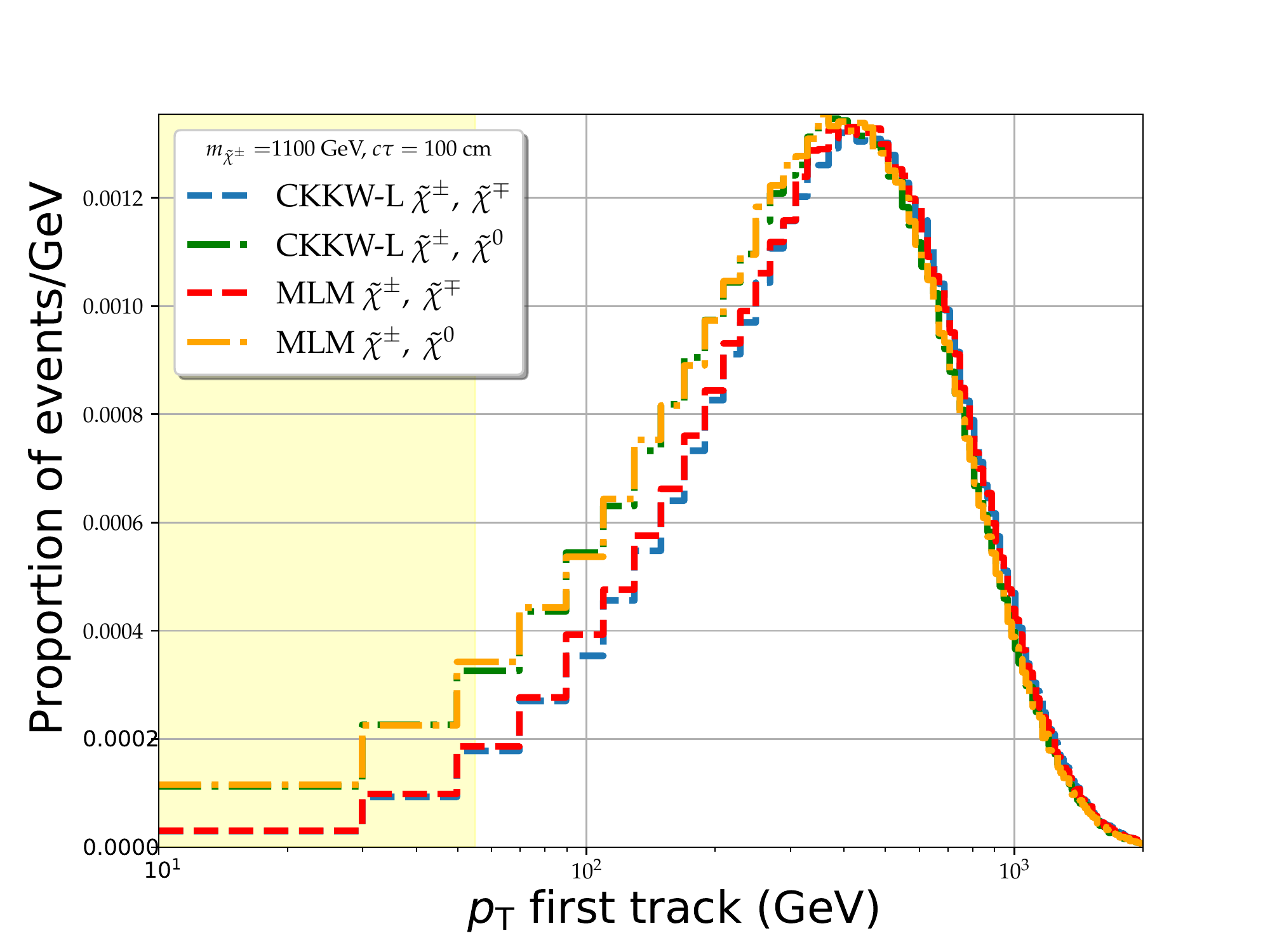}
  \caption{\label{FIG:pT}Comparison of the distribution of highest $p_T$ chargino per event (labelled as $p_T$ first track vs proportion of events per GeV in each bin) for double-chargino and single-chargino events, for both CKKW-L merging and MLM matching. Left: $100$ GeV winos; right: $1100$ GeV winos. The distribution uses particle-level information before any detector considerations, so is independent of decay length. Not shown is the overflow bin, which contains negligible events.}
\end{figure}

The effect of merging/matching on the distribution of $p_T$ of the charginos is shown in figure \ref{FIG:pT}, for $100$ GeV winos and $1100$ GeV winos. As in the MET case, both distributions are smooth and there is no obvious superior choice just from examining the plots, but substantial differences between the two can be seen in particular for lighter winos, which again contribute to the uncertainty. As a result, in the validation of the code we provide more than one set of results so that the effect of matching/merging can be seen, which yields differences up to about $30\%$ overall.

\section{Validation}
\label{SEC:VALIDATION}

The recasting material provided in \HEPData includes cutflows for six benchmark points ($300$ GeV and $700$ GeV, lifetimes of 10cm, 100cm and 1000cm for each) for both the wino and higgsino cases; acceptances for each signal region for masses between 100 \GeV and 1100 \GeV and a large range of lifetimes; and of course the exclusion plots. There is not a significant difference between the wino and higgsino cases in terms of cutflows; the main difference is the ratio of the two-charged-track events to single charged tracks, which for the wino case is roughly $1:2$ and for the higgsino case $2:7$. In general, the efficiencies for two-charged-track events are roughly twice those for the single-charge-track events, with a reduction at longer lifetimes because the presence of a very long track interferes with the MET calculation.

There is therefore in principle a large amount of data for validation, in fact too much to present in it entirety here. The classic standard is to compare cutflows, and we present a selection of these in appendix \ref{APP:CUTFLOWS}, with many more available online at the address given in sec.~\ref{SEC:HACKANALYSIS}. We provide a comparison of the three different matching/merging approaches in table \ref{TAB:MERGING}, where it can be seen that the \MG MLM and CKKW-L approaches give the best agreement for the chosen data point. For each point we simulated 500k events which, when split across 15 cores on a 40-core 3500 GHz cluster computer, took about an hour per point to simulate \emph{both} MLM and CKKW-L events combined. In table \ref{TAB:PILEUP} we demonstrate the (marginal) effect of the pileup on the cutflow, as discussed previously. In tables \ref{TAB:10cm}, \ref{TAB:100cm} and \ref{TAB:1000cm} we compare all of the the benchmark points for one signal region, namely 2018A, for the \MG MLM matching (which uses reweighting rather than vetos) which we denote ``HEPMC,'' since that is the input mode for the code \HA. Throughout good agreement can be seen.

While we have discussed the uncertainties coming from the different matching/merging approaches on the theoretical side, the cutflows -- and especially the acceptances provided by CMS on \HEPData -- actually come with substantial uncertainties too. The reason for this is the enormous statistics required: the efficiencies for the signal regions in this analysis range from $10^{-2}$ down to $10^{-5}$ (or lower) with most values being from $10^{-4}$ to $10^{-3}$; in order to have a $10\%$ uncertainty on an efficiency of $10^{-4}$ one would have to simulate $10^6$ events, whereas from the quoted statistical errors it appears that CMS simulated of the order of $5\times 10^4$ for each cutflow table and $3 \times 10^3$ for each point in the acceptances table. Moreover, while the acceptances are quoted for a very large number of lifetimes, events were only simulated for a small number and the results for intermediate lifetimes are computed by reweighting the events according to the decay length. These are from $0.2$cm to $10000$cm in logarithmic steps. For example, let us look at the quoted acceptances for single chargino events (which are the most important) and take the best case at short lifetimes: a $100$ GeV wino. The acceptances for signal region 2018A are:
\begin{equation}
  \begin{array}{|c|c|}  \hline\hline
    \mathrm{Decay\ length\ (cm)} & \mathrm{Acceptance} \\ \hline
    0.2 & 6.4\times10^{-13} \pm 6\times 10^{-13} \\
    1 &  1.15\times 10^{-5} \pm 7.4 \times 10^{-6} \\
    2 & 9.5 \times 10^{-5} \pm 4.0 \times 10^{-5} \\
    10 & 5.6 \times 10^{-4} \pm 1.5 \times 10^{-4} \\ \hline
    100 & 2.4 \times 10^{-4} \pm 1.0 \times 10^{-4} \\
    1000 & 2.5 \times 10^{-5} \pm 2.5 \times 10^{-5}\\
    10000 & 4.3 \times 10^{-6} \pm 4.3 \times 10^{-6}\\\hline\hline
   \end{array}
\end{equation}
The data up to decay lengths of about $10$cm is therefore up to $100$\% uncertainty and there is little point attempting to match to better than a factor of $2$ (or even at all below $1$cm). Similarly at longer decay lengths a similar story plays out for this signal region (i.e. the data is not meaningful). Instead it is only useful to compare to the ``best'' number of layers for any point, as that is where the statistical power is greatest. Therefore to give a reasonable measure of the uncertainty from our code we compute
\begin{align}
\mathrm{\delta} \equiv&   \frac{\sum_i  (\epsilon_i^{CMS} - \epsilon_i) \mathcal{L}_i}{ \sum_i \epsilon_i^{CMS}\mathcal{L}_i} \qquad i \in \{2015,2016A,2016B, 2017,2018A,2018B\}
\end{align}
where $\epsilon_i^{CMS}$ is the acceptance according to \HEPData for the \emph{best} signal region (i.e. the number of layers that produces the best \emph{expected} limit according to the $\mathrm{CL}_s$ procedure, where by expected we mean taking the number of observed events equal to the background) for the data-taking period $i$; $\epsilon_i$ is the same thing for our code; and $\mathcal{L}_i$ is the integrated luminosity of the period $i$. In effect we are comparing the predicted total number of events over the whole $139\ \invfb$, effectively including the 2015 and 2016 data in the $n_{\rm lay} \ge 6$ region. The values for single chargino events are given in table \ref{TAB:AcceptanceCHNU} and for double chargino events in table \ref{TAB:AcceptanceCHCH}. We do not show the values where the uncertainty on the acceptances is too high (i.e. for some $100$ GeV values). As can be seen, good agreement is found over the whole range for both MLM and CKKW-L matching/merging, with perhaps better agreement for the latter. Especially given the above observations about the uncertainties in the \HEPData, and the fact that the acceptance for this analysis is a very rapidly changing function of both mass and decay length, the accuracy is entirely adequate, as will be seen in the next section.

\begin{table} \centering 
  \begin{tabular}{||c||c|c|c|c|c|c|c|c|c|c|c||} \hline \hline
    \multicolumn{12}{||c||}{CKKW-L merging} \\ \hline
 & \multicolumn{11}{c||}{Mass (GeV)} \\ 
 Decay Length (cm) 
 & 100 & 200 & 300 & 400 & 500 & 600 & 700 & 800 & 900 & 1000 & 1100 \\ \hline
 10  & --  & 52\%  & 14\%  & 6\%  & 5\%  & 13\%  & 8\%  & 3\%  & -2\%  & -20\%  & -23\% \\ \hline 
 100  & --  & 35\%  & 40\%  & 30\%  & 10\%  & 4\%  & 25\%  & 10\%  & 10\%  & -18\%  & -12\% \\ \hline 
 1000  & --  & 47\%  & 33\%  & 25\%  & 47\%  & 20\%  & 24\%  & 30\%  & 31\%  & -19\%  & -48\% \\ \hline 
 10000  & 25\%  & 16\%  & 21\%  & 6\%  & -18\%  & -30\%  & -36\%  & -23\%  & -36\%  & -47\%  & -1\% \\ \hline 
\hline 
    \multicolumn{12}{||c||}{MLM matching} \\ \hline
    & \multicolumn{11}{c||}{Mass (GeV)} \\ 
 Decay Length (cm) 
 & 100 & 200 & 300 & 400 & 500 & 600 & 700 & 800 & 900 & 1000 & 1100 \\ \hline
 10  & --  & 56\%  & 36\%  & 21\%  & 25\%  & 31\%  & 29\%  & 25\%  & 19\%  & 2\%  & 3\% \\ \hline 
 100  & --  & 30\%  & 46\%  & 47\%  & 33\%  & 24\%  & 39\%  & 31\%  & 27\%  & 4\%  & 9\% \\ \hline 
 1000  & --  & 57\%  & 42\%  & 39\%  & 56\%  & 35\%  & 35\%  & 46\%  & 50\%  & 16\%  & 26\% \\ \hline 
 10000  & 34\%  & 29\%  & 43\%  & 31\%  & 7\%  & -10\%  & 2\%  & -5\%  & -6\%  & -20\%  & 22\% \\ \hline 
\hline  
\end{tabular} 
\caption{\label{TAB:AcceptanceCHNU}Errors for chargino-neutralino events. Top: using CKKW-L merging; Bottom: using MLM matching.} 
\end{table} 
\begin{table} \centering 
  \begin{tabular}{||c||c|c|c|c|c|c|c|c|c|c|c||} \hline \hline
      \multicolumn{12}{||c||}{CKKW-L merging} \\ \hline
 & \multicolumn{11}{c||}{Mass (GeV)} \\ 
 Decay Length (cm) 
 & 100 & 200 & 300 & 400 & 500 & 600 & 700 & 800 & 900 & 1000 & 1100 \\ \hline
 10  & 52\%  & 42\%  & 32\%  & 19\%  & 13\%  & 13\%  & 6\%  & 3\%  & -4\%  & -30\%  & -25\% \\ \hline 
 100  & --  & 53\%  & 35\%  & 14\%  & 16\%  & 17\%  & 13\%  & 6\%  & -13\%  & -10\%  & -11\% \\ \hline 
 1000  & 59\%  & 17\%  & 48\%  & 50\%  & 32\%  & 26\%  & 19\%  & 16\%  & 10\%  & -40\%  & -12\% \\ \hline 
 10000  & 30\%  & 20\%  & 12\%  & 5\%  & -66\%  & -14\%  & -57\%  & -29\%  & -31\%  & -49\%  & -25\% \\ \hline 
    \hline
  \multicolumn{12}{||c||}{MLM matching} \\ \hline   
 & \multicolumn{11}{c||}{Mass (GeV)} \\ 
 Decay Length (cm) 
 & 100 & 200 & 300 & 400 & 500 & 600 & 700 & 800 & 900 & 1000 & 1100 \\ \hline
 10  & 46\%  & 47\%  & 42\%  & 32\%  & 28\%  & 27\%  & 24\%  & 21\%  & 15\%  & -7\%  & -5\% \\ \hline 
 100  & --  & 50\%  & 45\%  & 31\%  & 33\%  & 29\%  & 28\%  & 23\%  & 11\%  & 9\%  & 6\% \\ \hline 
 1000  & 92\%  & 30\%  & 65\%  & 51\%  & 33\%  & 35\%  & 31\%  & 37\%  & 31\%  & -28\%  & -15\% \\ \hline 
 10000  & 33\%  & 34\%  & 26\%  & 19\%  & -34\%  & 1\%  & -29\%  & -1\%  & -12\%  & -19\%  & 2\% \\ \hline \hline  
\end{tabular} 
\caption{\label{TAB:AcceptanceCHCH}Errors for chargino-chargino events. Top: using CKKW-L merging; Bottom: using MLM matching.} 
\end{table}

\section{Long dead winos}
\label{SEC:WINOS}

Having validated our code by comparing with cutflows and acceptances, here we reproduce the exclusion limits for winos, which is a classic $SU(2)$ triplet fermion having zero hypercharge and can be regarded as a Minimal Dark Matter candidate. %
However, since we have a recasting code, we can do more: we can also apply the constraints from other analyses. In particular, very long-lived winos can also be looked for in searches for heavy stable charged particles; there is the analysis \cite{Aaboud:2019trc} by ATLAS which provided extensive recasting material including efficiencies and pseudocode. Notably this formed the basis for a code on the {\tt LLPrecasting} repository \url{https://github.com/llprecasting/recastingCodes/blob/master/HSCPs/ATLAS-SUSY-2016-32/}. By treating any chargino that escapes the muon system as a stable particle, and approximating the ATLAS muon chambers as a cylinder of radius 12m and of length 46m (so $|z| < 23$m) we can use this analysis to constrain longer-lived charginos. Although the ATLAS analysis has only $36.1 \invfb$ it is especially powerful, since a heavy particle in the muon system is a rather striking signal, and so together these two anaylyses can provide overlapping regions of exclusion.

For the CMS disappearing track analysis, we generate two charged track and one charged track events separately, and use the same data as for tables \ref{TAB:AcceptanceCHCH} and \ref{TAB:AcceptanceCHNU} supplemented by additional refinement points at intermediate lifetimes. To calculate exclusion limits we use a python code (available online) to combine the results from different data-taking periods in the same signal region (and treat the 2015/2016 data as belonging to signal region 3) to produce a $\mathrm{CL}_s$ value, and exclude points at $95\%$ confidence level. For the production cross-sections we use the publicly-available NLO-NLL results \cite{Fuks:2012qx,Fuks:2013vua} from \url{https://twiki.cern.ch/twiki/bin/view/LHCPhysics/SUSYCrossSections}.

For the ATLAS heavy stable charged particle analysis, we used the leading-order \pythia code from the {\tt LLPrecasting} repository, with the same cross-sections and a bespoke code to calculate the exclusion limits. 

\begin{figure}\centering
 \includegraphics[width=0.7\textwidth]{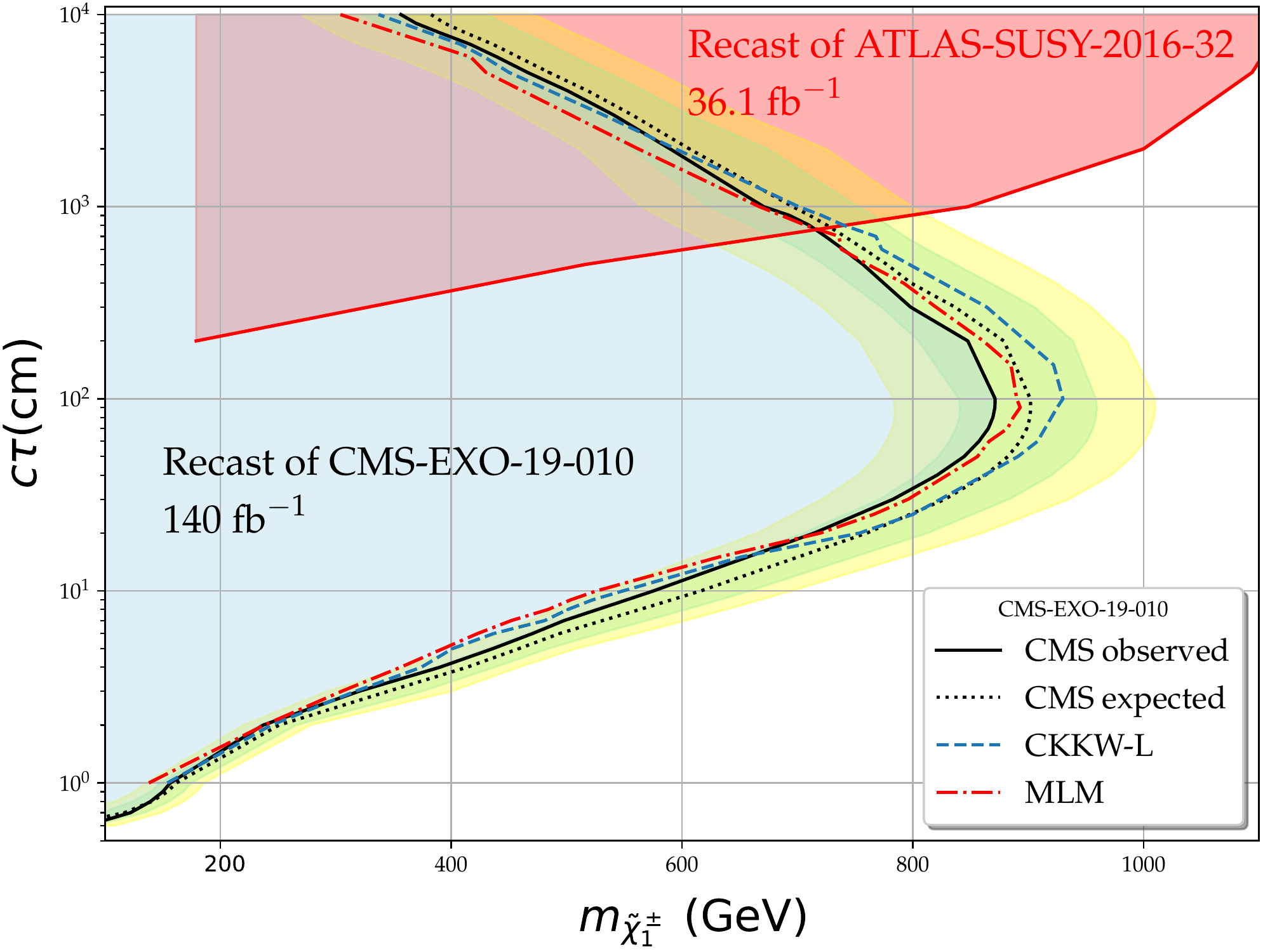}
\caption{\label{FIG:winocombo} Limits on winos from both the CMS disappearing track search and ATLAS heavy stable charged particle search.}
\end{figure}

The results are shown in figure \ref{FIG:winocombo}. As can be seen the excluded regions overlap so that $200$ GeV winos are excluded for \emph{any} decay lengths ($c\tau$) longer than about 2cm, and $700$ GeV winos for lengths longer than about $20$cm. One peculiarity of the ATLAS search is that it is only sensitive to masses above around $175$ GeV (since the mass measurement relies on time of flight information they placed a cut on the lowest masses). Hence in principle the disappearing track search is most sensitive for very light winos even at rather long lifetimes, and a (meta-) stable wino below about $175$ GeV cannot be ruled out by these searches.    

Most importantly, however, we see that our approach is well able to reproduce the exclusion plot for both MLM matching and CKKW-L merging approaches, and the differences between them give a measure of the uncertainty in the result. 

As a final caveat on these results, we note that for very long-lived charginos -- or, indeed, any SU(2) multiplet with a (meta-)stable neutral component -- the production of a neutral and charged fermion together leads to very large missing energy (as can be seen from the differences in fig.~\ref{FIG:MET700}, which should be a trigger for prompt searches and may lead to additional exclusions (which are already excluded by our results in this case). It would be interesting to explore this in the context of other models, but would require a recasting of the relevant analyses to take this effect into account and we leave it for future work.

\section{Limits on light charged scalars}
\label{SEC:SCALARS}

Having demonstrated the versatility of the code and reproduced the CMS data, here we apply it to a model with charged scalars. If we add a single charged scalar $S^-$ with hypercharge $1$ (and charged under lepton number) and a neutral fermion $\tilde{\chi}^0$ then the most general Lagrangian is
\begin{align}
\mathcal{L} = \mathcal{L}_{SM} - m_{S^\pm}^2 |S^-|^2 - \frac{1}{4}\lambda_2 |S^-|^4- \lambda_3 |S^-|^2 |H|^2 - \bigg[ y_{RS} \ov{S}^- e^c \tilde{\chi}^0  - \frac{1}{2} m_{\tilde{\chi}^0} \tilde{\chi}^0 \tilde{\chi}^0 + h.c. \bigg].
\end{align}
Clearly this is a prototype of a bino and right-handed slepton in supersymmetry, except that the couplings  $\lambda_2, \lambda_3$ and $y_{RS}$  are undetermined. This is an excellent  (if rather fine-tuned) dark matter model, and the classic LHC constraints would be dominated by pair production of $S^\pm$ via a Z boson; the cross-sections are small but there should be limits from conventional monojet/monophoton (mono-X) searches. As discussed in \cite{Belyaev:2020wok}, such searches are very generic, relying only that new hidden particles are produced so that they give missing transverse energy to recoil against, but in the absence of new heavy mediators that can be produced on-shell give only very weak limits; for this model the current limits should not be significant, although it would be interesting to explore them for the HL-LHC. Moreoever, in contrast to the wino/SU(2) multiplet case, since the charged $S^\pm$ bosons are only produced in pairs, long-lived scalars do not generally lead to large missing energy since they will either both be classed as muons or neither (if the decay length is short enough).

\begin{figure}\centering
 \includegraphics[width=0.6\textwidth]{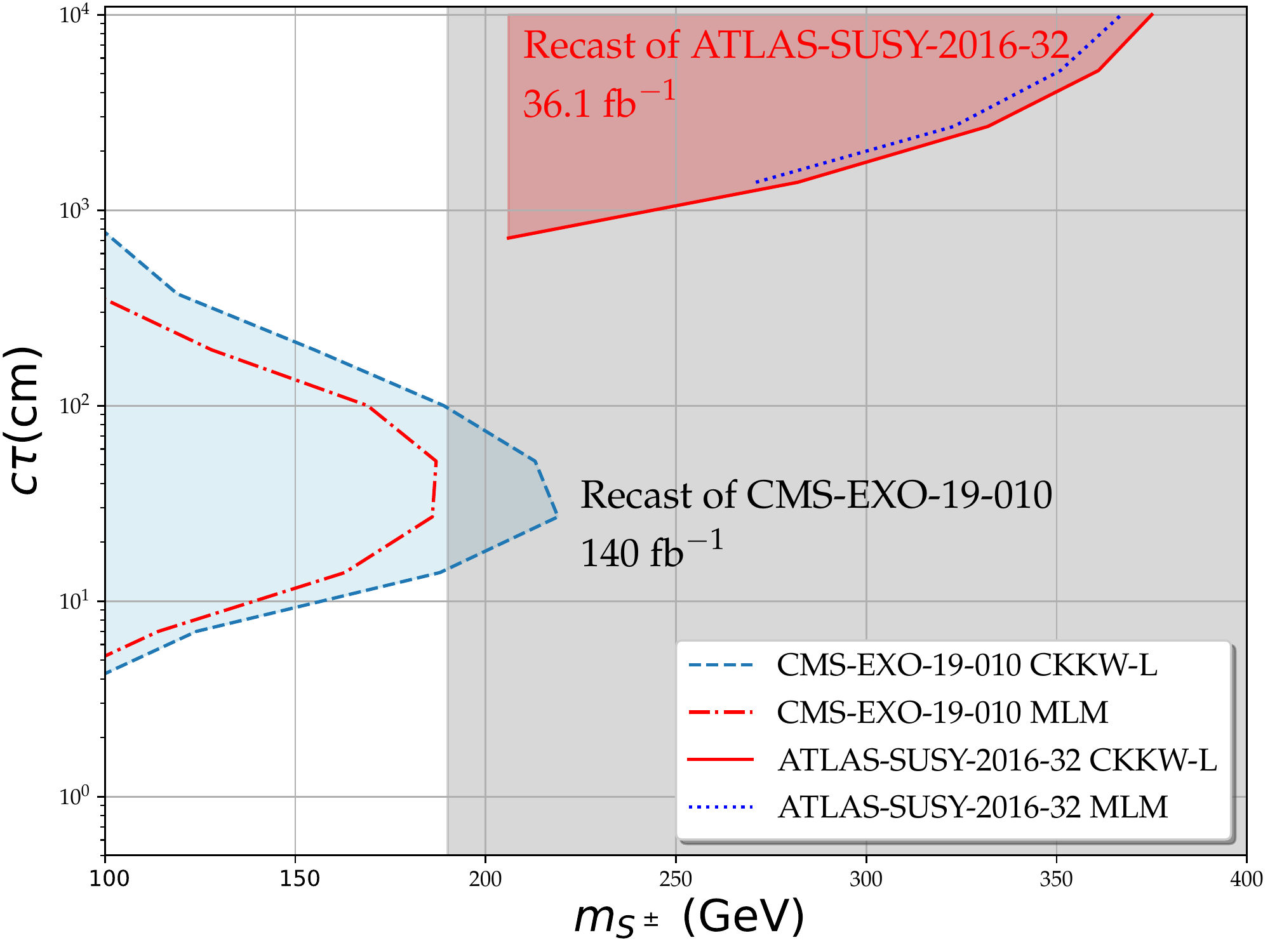}
\caption{\label{FIG:scalarcombo} Limits on charged scalar particles. The excluded regions from the disappearing track and heavy stable charged particle searches are shown; the dark grey region is excluded as giving an overdense dark matter relic abundance.}
\end{figure}

The width of the charged scalar decay (neglecting the electron mass) is given by
\begin{align}
  \Gamma (\phi \rightarrow \chi e) \simeq& \frac{1}{16\pi m_{S^\pm}^3} (m_{S^\pm}^2 - m_{\tilde{\chi}^0}^2)^2 |y_{RS}|^2 \simeq \frac{1}{4\pi m_{S^\pm}} (m_{S^\pm} - m_{\tilde{\chi}^0})^2 |y_{RS}|^2 
\end{align}
so we can have a long lifetime for the charged scalar via a small mass difference and/or coupling. We fix $m_{S^\pm} - m_{\tilde{\chi}^0} = 90\ \mathrm{MeV} < m_\pi $ (so that even loop-induced decays to pions are impossible, since we want to consider a different channel to the wino model) so that the proper decay length is
\begin{align}
c\tau \simeq& 100\ \mathrm{cm} \times \left(\frac{m_{S^\pm}}{100\ \mathrm{GeV}}\right) \times \left( \frac{ 90\ \mathrm{MeV}}{m_{S^\pm} - m_{\tilde{\chi}^0}} \right)^2 \times \left( \frac{5.5 \times 10^{-6}}{y_{RS}} \right)^2 .
\end{align}
We implemented this model in \SARAH \cite{Staub:2008uz,Staub:2009bi,Staub:2010jh,Staub:2012pb,Staub:2013tta,Goodsell:2017pdq} to calculate the spectra and decays precisely, and produced a {\tt UFO} \cite{Degrande:2011ua} for \MG to generate events for the LHC searches. We computed the limits from disappearing tracks and heavy stable charged particle searches using \HA, and, since this can be a dark matter model, we also computed the dark matter relic density using {\tt MicrOMEGAs5.2} \cite{Belanger:2018ccd,Belanger:2020gnr}. The results are shown in figure \ref{FIG:scalarcombo}. In contrast to the wino case, the HSCP and DT search exclusion regions do not overlap, because the DT limits are much weaker thanks to the small cross-section. On the other hand, the HSCP region is excluded by the dark matter density, while there is some complementarity between the DT and DM searches. The parameter space is then viable for smaller masses and both smaller and longer lifetimes.

\section{A hackable recasting code}
\label{SEC:HACKANALYSIS}

Here we describe the code that incorporates the recast of the CMS disappearing track search, and also another version of the ATLAS heavy stable particle search, so that both can be run on the same simulated events. This code is called \HA and is available at \url{https://github.com/llprecasting/recastingCodes/tree/master/DisappearingTracks/CMS-EXO-19-010}; it is maintained at \url{https://goodsell.pages.in2p3.fr/hackanalysis} where versions with any other analyses in future will appear.

As mentioned in the introduction, \HA is not meant to be a new framework, but merely to be used for prototyping so that analyses/features can be exported e.g. to \MA. It is therefore designed to be flexible and editable to give the user complete control at every stage without obfuscation. It has three modes of operation: (1) event generation using \pythia; (2) reading {\tt LesHouchesEvent} ({\tt .lhe}) files (presumably from \MG) and showering through \pythia; (3) reading \hepmcV files. In the latter two cases, either compressed (via {\tt gzip}) or uncompressed files are accepted. In modes (1) and (2) multicore operation is possible (via {\tt pragma omp}); in mode (2) this means the {\tt .lhe} or {\tt .lhe.gz} files should be split into one file per core (which can be performed automatically with a \python script provided).

As a basis for the event handling, events from \pythia or \hepmc are converted to a common event format which is based on a modified version of the {\tt heputils} (\url{https://gitlab.com/hepcedar/heputils}) package where, in compliance with the licence, the namespace is renamed (as {\tt HEP}). In addition some code is taken from {\tt mcutils}  and in principle smearing can be applied identically to GAMBIT although this is not actually used for these analyses. 

The code \HA is very lightweight, making use as far as possible of existing libraries; this means that editing and (re)building is very fast. It therefore requires \pythia \cite{Sjostrand:2014zea} (from version $8.303$) for showering/event generation; \fastjet \cite{Cacciari:2011ma} for jet clustering (in principle this could be slimmed down to {\tt fjcore}, but it may be desirable to use the more advanced features for e.g. pileup subtraction); \YODA (\url{https://yoda.hepforge.org/}) to read {\tt YAML} files and handle histogramming (and the reading of \HEPData efficiency tables if required, although so far no analyses rely on this feature) and \hepmcV if reading/writing that format is desired. Compilation is straightforward on a unix-based system; the user must only provide the paths for the necessary packages in the {\tt Makefile}:
\begin{lstlisting}
hepmcpath=<HepMC2 installation directory>
pythia8path=<pythia directory>
fastjetpath=<fastjet installation directory>

YODApath=<YODA installation directory>
YAMLpath=<YODA code directory>/src/yamlcpp
\end{lstlisting}
The only subtlety here is the {\tt YAMLpath}; the code uses the YAML reader included in YODA, but a peculiarity of that package is that the header files are not installed in the installation directory, so this path must point to the directory where the YODA code is stored. Note that it would be straightforward to include a separate YAML reader at the expense of installing more packages. 

Once the code is built, a library and three executables are created, named {\tt analysePYTHIA.exe},\\ {\tt analysePYTHIA\_LHE.exe} and {\tt analyseHEPMC.exe}, corresponding to modes (1), (2) and (3) respectively. All three accept a {\tt YAML} file to specify settings such as which analyses to run, the names of \pythia configuration files, names of output files, whether to include pileup, etc.

An example {\tt YAML} file to run the program would be:
\begin{lstlisting}
---
analyses:
 - DT_CMS
 - HSCP_ATLAS
 
settings:
  nevents: 50000
  cores: 4
  Config file: pythia_config.cfg
  LHE file: MGdir/unweighted_events
  HEPMC file: unweighted_events.hepmc.gz
  Efficiency Filename: output_efficiencies.eff
  Cutflow Filename: output_cutflow.txt
  Histogram Filename: histos.yoda
  Include Pileup: true
  Pileup Filename: minbias.tar.gz
  Average Pileup: 29.0
\end{lstlisting}

The output is a set of text files: an efficiency file (which contains the efficiency and uncertainty of each signal region) which resembles an SLHA format; a cutflow file (which prints the cutflows in a verbose text form); and a file of \YODA histograms. Example input files, configuration files for \pythia for modes (1) and (2) (for MLM/CKKW-L matching/merging) and python files for reading the output and calculating exclusion limits for the included analyses are provided. Moreover, code for generating and storing a pileup event file are provided.

\section{Conclusions}
\label{SEC:CONCLUSIONS}

We have presented an approach and a code to recast the CMS disappearing track searches using the full run 2 data, and shown that in combination with heavy stable charged particle searches a large part of the parameter space can be excluded for a wide variety of models. We have presented extensive validation material and discussion of the technical challenges, as well as some exclusions for a new model. 

It would now be interesting to apply the code to produce tables of efficiencies for each signal region for different classes of models (scalars with one or two tracks, vectors with one or two tracks) so that searches can be recast in a simplified models approach. Clearly, with three signal regions and six data-taking periods it is impractical to publish such tables in a paper, but they should ultimately be available online. It would also be useful to apply the results to complete models such as those in \cite{Das:2020uer} or  \cite{Goodsell:2020lpx} and explore new scenarios. We hope to return to these in future work. Moreover, the analysis described here should also be made available in other frameworks, in particular \MA where work to do so is underway.

\section*{Acknowledgments}

We thank Brian Francis for very helpful discussions about the CMS analysis. We thank Sabine Kraml, Humberto Reyes Gonzalez and Sophie Williamson for collaboration on related topics; Andre Lessa for helpful discussions; and Jack Araz, Benjamin Fuks, Manuel Utsch for collaboration on LLP recasting in \MA, and Benjamin Fuks for comments on the draft. We thank the organisers of the LLP workshops \url{https://longlivedparticles.web.cern.ch/}, especially 
MDG acknowledges support from the grant
\mbox{``HiggsAutomator''} of the Agence Nationale de la Recherche
(ANR) (ANR-15-CE31-0002). 

\appendix

 \section{Cutflow comparisons}
\label{APP:CUTFLOWS}

\begin{table}\centering
\begin{tabular}{|c||c|c|c||} \hline\hline
 & \multicolumn{3}{|c||}{700 GeV, 10 cm, region 2018A} \\ Cut & $\epsilon_i^{\rm CMS}$ & $\epsilon_i^{\rm sim}$, HEPMC &$\epsilon_i^{\rm sim}$, HEPMC, no pileup \\\hline \hline
{\footnotesize total                                                                          } &  $ 1.0^{+0.00}_{-0.00} $  & $ 1.0^{+0.00}_{-0.00} $  &  $ 1.0^{+0.00}_{-0.00} $  \\
{\footnotesize trigger                                                                        } &  $ 1.5^{+0.02}_{-0.02}\times10^{-1} $  & $ 1.5^{+0.01}_{-0.01}\times10^{-1} $  &  $ 1.5^{+0.01}_{-0.01}\times10^{-1} $  \\
{\footnotesize passes $p_{\text{T}}^{\text{miss}}$ filters                                    } &  $ 1.4^{+0.02}_{-0.02}\times10^{-1} $  & $ 1.5^{+0.01}_{-0.01}\times10^{-1} $  &  $ 1.5^{+0.01}_{-0.01}\times10^{-1} $  \\
{\footnotesize $p_{\text{T}}^{\text{miss}} > 120\,\text{GeV}{}$                               } &  $ 1.4^{+0.02}_{-0.02}\times10^{-1} $  & $ 1.5^{+0.01}_{-0.01}\times10^{-1} $  &  $ 1.5^{+0.01}_{-0.01}\times10^{-1} $  \\
{\footnotesize $\geq 1$ jet with $p_{\text{T}} > 110\,\text{GeV}{}$ and $|\eta| < 2.4$        } &  $ 1.3^{+0.02}_{-0.02}\times10^{-1} $  & $ 1.3^{+0.01}_{-0.01}\times10^{-1} $  &  $ 1.3^{+0.01}_{-0.01}\times10^{-1} $  \\
{\footnotesize ==0 pairs of jets with $\Delta\phi_{\text{jet, jet}} > 2.5$                    } &  $ 1.1^{+0.01}_{-0.01}\times10^{-1} $  & $ 1.1^{+0.01}_{-0.01}\times10^{-1} $  &  $ 1.1^{+0.01}_{-0.01}\times10^{-1} $  \\
{\footnotesize $|\Delta\phi(\mbox{leading jet}, \vec{p}_{\text{T}}^{\text{miss}})| > 0.5$     } &  $ 1.1^{+0.01}_{-0.01}\times10^{-1} $  & $ 1.1^{+0.01}_{-0.01}\times10^{-1} $  &  $ 1.1^{+0.01}_{-0.01}\times10^{-1} $  \\
{\footnotesize $\geq 1$ track with $|\eta| < 2.1$                                             } &  $ 1.1^{+0.01}_{-0.01}\times10^{-1} $  & $ 1.1^{+0.01}_{-0.01}\times10^{-1} $  &  $ 1.1^{+0.01}_{-0.01}\times10^{-1} $  \\
{\footnotesize $\geq 1$ track with $p_{\text{T}} > 55\,\text{GeV}{}$                          } &  $ 4.7^{+0.10}_{-0.10}\times10^{-2} $  & $ 4.6^{+0.06}_{-0.06}\times10^{-2} $  &  $ 4.6^{+0.06}_{-0.06}\times10^{-2} $  \\
{\footnotesize $\geq 1$ track passing fiducial selections                                     } &  $ 3.1^{+0.08}_{-0.08}\times10^{-2} $  & $ 3.6^{+0.05}_{-0.05}\times10^{-2} $  &  $ 3.6^{+0.05}_{-0.05}\times10^{-2} $  \\
{\footnotesize $\geq 1$ track with $\geq 4$ pixel hits                                        } &  $ 1.7^{+0.06}_{-0.06}\times10^{-2} $  & $ 2.7^{+0.05}_{-0.05}\times10^{-2} $  &  $ 2.7^{+0.05}_{-0.05}\times10^{-2} $  \\
{\footnotesize $\geq 1$ track with no missing inner hits                                      } &  $ 1.7^{+0.06}_{-0.06}\times10^{-2} $  & $ 2.0^{+0.04}_{-0.04}\times10^{-2} $  &  $ 2.0^{+0.04}_{-0.04}\times10^{-2} $  \\
{\footnotesize $\geq 1$ track with no missing middle hits                                     } &  $ 1.5^{+0.05}_{-0.05}\times10^{-2} $  & $ 2.0^{+0.04}_{-0.04}\times10^{-2} $  &  $ 2.0^{+0.04}_{-0.04}\times10^{-2} $  \\
{\footnotesize $\geq 1$ track with relative track isolation $< 5\%$                           } &  $ 5.3^{+0.34}_{-0.34}\times10^{-3} $  & $ 6.0^{+0.23}_{-0.23}\times10^{-3} $  &  $ 6.2^{+0.23}_{-0.23}\times10^{-3} $  \\
{\footnotesize $\geq 1$ track with $|d_{\text{xy}}| < 0.02\,\text{cm}{}$                      } &  $ 5.1^{+0.34}_{-0.34}\times10^{-3} $  & $ 6.0^{+0.23}_{-0.23}\times10^{-3} $  &  $ 6.2^{+0.23}_{-0.23}\times10^{-3} $  \\
{\footnotesize $\geq 1$ track with $|d_z| < 0.5\,\text{cm}{}$                                 } &  $ 5.1^{+0.31}_{-0.31}\times10^{-3} $  & $ 6.0^{+0.23}_{-0.23}\times10^{-3} $  &  $ 6.2^{+0.23}_{-0.23}\times10^{-3} $  \\
{\footnotesize $\geq 1$ track with $\Delta R (\text{track}, \text{jet}) > 0.5$                } &  $ 4.9^{+0.31}_{-0.31}\times10^{-3} $  & $ 5.9^{+0.22}_{-0.22}\times10^{-3} $  &  $ 6.1^{+0.23}_{-0.23}\times10^{-3} $  \\
{\footnotesize $\geq 1$ track with $\Delta R (\text{track}, \text{electron}) > 0.15$          } &  $ 4.7^{+0.31}_{-0.31}\times10^{-3} $  & $ 5.9^{+0.22}_{-0.22}\times10^{-3} $  &  $ 6.1^{+0.23}_{-0.23}\times10^{-3} $  \\
{\footnotesize $\geq 1$ track with $\Delta R (\text{track}, \text{muon}) > 0.15$              } &  $ 4.7^{+0.31}_{-0.31}\times10^{-3} $  & $ 5.9^{+0.22}_{-0.22}\times10^{-3} $  &  $ 6.1^{+0.23}_{-0.23}\times10^{-3} $  \\
{\footnotesize $\geq 1$ track with $\Delta R (\text{track}, \tau_{\text{h}}) > 0.15$          } &  $ 4.7^{+0.31}_{-0.31}\times10^{-3} $  & $ 5.9^{+0.22}_{-0.22}\times10^{-3} $  &  $ 6.1^{+0.23}_{-0.23}\times10^{-3} $  \\
{\footnotesize $\geq 1$ track with $E_{\text{calo}} < 10\,\text{GeV}{}$                       } &  $ 4.7^{+0.31}_{-0.31}\times10^{-3} $  & $ 5.9^{+0.22}_{-0.22}\times10^{-3} $  &  $ 6.1^{+0.23}_{-0.23}\times10^{-3} $  \\
{\footnotesize $\geq 1$ track with $\geq 3$ missing outer hits                                } &  $ 4.6^{+0.31}_{-0.31}\times10^{-3} $  & $ 5.9^{+0.22}_{-0.22}\times10^{-3} $  &  $ 6.0^{+0.23}_{-0.23}\times10^{-3} $  \\
{\footnotesize $\geq 1$ track $4$ layers                                                      } &  $ 2.8^{+0.24}_{-0.24}\times10^{-3} $  & $ 3.2^{+0.17}_{-0.17}\times10^{-3} $  &  $ 3.4^{+0.17}_{-0.17}\times10^{-3} $  \\
{\footnotesize $\geq 1$ track $5$ layers                                                      } &  $ 9.2^{+1.36}_{-1.36}\times10^{-4} $  & $ 1.4^{+0.11}_{-0.11}\times10^{-3} $  &  $ 1.4^{+0.11}_{-0.11}\times10^{-3} $  \\
{\footnotesize $\geq 1$ track $\geq 6$ layers                                                 } &  $ 9.5^{+1.36}_{-1.36}\times10^{-4} $  & $ 1.2^{+0.10}_{-0.10}\times10^{-3} $  &  $ 1.2^{+0.10}_{-0.10}\times10^{-3} $  \\
\hline\hline\end{tabular}
\caption{\label{TAB:PILEUP}Comparison of the effect of pileup on the cutflows for 700 GeV, 10 cm winos, region 2018A. ``HEPMC'' denotes processing of events stored in \hepmc files produced by \MG.}
\end{table}

\begin{table}\centering
\begin{tabular}{|c||c|c|c|c||} \hline\hline
 & \multicolumn{4}{|c||}{700 GeV, 1000 cm, region 2017} \\ Cut & $\epsilon_i^{\rm CMS}$ & $\epsilon_i^{\rm sim}$, HEPMC &$\epsilon_i^{\rm sim}$, CKKW-L & $\epsilon_i^{\rm sim}$, MLM  \\\hline \hline
{\footnotesize total                                                                          } &  $ 1.0^{+0.00}_{-0.00} $  & $ 1.0^{+0.00}_{-0.00} $  &  $ 1.0^{+0.00}_{-0.00} $  & $ 1.0^{+0.00}_{-0.00} $ \\
{\footnotesize trigger                                                                        } &  $ 2.0^{+0.02}_{-0.02}\times10^{-1} $  & $ 1.8^{+0.01}_{-0.01}\times10^{-1} $  &  $ 2.1^{+0.01}_{-0.01}\times10^{-1} $  & $ 1.7^{+0.01}_{-0.01}\times10^{-1} $ \\
{\footnotesize passes $p_{\text{T}}^{\text{miss}}$ filters                                    } &  $ 2.0^{+0.02}_{-0.02}\times10^{-1} $  & $ 1.8^{+0.01}_{-0.01}\times10^{-1} $  &  $ 2.1^{+0.01}_{-0.01}\times10^{-1} $  & $ 1.7^{+0.01}_{-0.01}\times10^{-1} $ \\
{\footnotesize $p_{\text{T}}^{\text{miss}} > 120\,\text{GeV}{}$                               } &  $ 1.9^{+0.02}_{-0.02}\times10^{-1} $  & $ 1.8^{+0.01}_{-0.01}\times10^{-1} $  &  $ 2.1^{+0.01}_{-0.01}\times10^{-1} $  & $ 1.7^{+0.01}_{-0.01}\times10^{-1} $ \\
{\footnotesize $\geq 1$ jet with $p_{\text{T}} > 110\,\text{GeV}{}$ and $|\eta| < 2.4$        } &  $ 1.4^{+0.02}_{-0.02}\times10^{-1} $  & $ 1.4^{+0.01}_{-0.01}\times10^{-1} $  &  $ 1.4^{+0.01}_{-0.01}\times10^{-1} $  & $ 1.0^{+0.01}_{-0.01}\times10^{-1} $ \\
{\footnotesize ==0 pairs of jets with $\Delta\phi_{\text{jet, jet}} > 2.5$                    } &  $ 1.2^{+0.02}_{-0.02}\times10^{-1} $  & $ 1.2^{+0.01}_{-0.01}\times10^{-1} $  &  $ 1.2^{+0.01}_{-0.01}\times10^{-1} $  & $ 9.3^{+0.08}_{-0.08}\times10^{-2} $ \\
{\footnotesize $|\Delta\phi(\mbox{leading jet}, \vec{p}_{\text{T}}^{\text{miss}})| > 0.5$     } &  $ 1.2^{+0.02}_{-0.02}\times10^{-1} $  & $ 1.2^{+0.01}_{-0.01}\times10^{-1} $  &  $ 1.1^{+0.01}_{-0.01}\times10^{-1} $  & $ 8.8^{+0.08}_{-0.08}\times10^{-2} $ \\
{\footnotesize $\geq 1$ track with $|\eta| < 2.1$                                             } &  $ 1.2^{+0.02}_{-0.02}\times10^{-1} $  & $ 1.2^{+0.01}_{-0.01}\times10^{-1} $  &  $ 1.1^{+0.01}_{-0.01}\times10^{-1} $  & $ 8.8^{+0.08}_{-0.08}\times10^{-2} $ \\
{\footnotesize $\geq 1$ track with $p_{\text{T}} > 55\,\text{GeV}{}$                          } &  $ 1.1^{+0.02}_{-0.02}\times10^{-1} $  & $ 1.1^{+0.01}_{-0.01}\times10^{-1} $  &  $ 1.0^{+0.01}_{-0.01}\times10^{-1} $  & $ 8.2^{+0.08}_{-0.08}\times10^{-2} $ \\
{\footnotesize $\geq 1$ track passing fiducial selections                                     } &  $ 7.9^{+0.12}_{-0.12}\times10^{-2} $  & $ 8.7^{+0.08}_{-0.08}\times10^{-2} $  &  $ 8.7^{+0.08}_{-0.08}\times10^{-2} $  & $ 6.9^{+0.07}_{-0.07}\times10^{-2} $ \\
{\footnotesize $\geq 1$ track with $\geq 4$ pixel hits                                        } &  $ 5.9^{+0.10}_{-0.10}\times10^{-2} $  & $ 7.0^{+0.07}_{-0.07}\times10^{-2} $  &  $ 6.9^{+0.07}_{-0.07}\times10^{-2} $  & $ 5.5^{+0.07}_{-0.07}\times10^{-2} $ \\
{\footnotesize $\geq 1$ track with no missing inner hits                                      } &  $ 5.9^{+0.10}_{-0.10}\times10^{-2} $  & $ 4.8^{+0.06}_{-0.06}\times10^{-2} $  &  $ 4.7^{+0.06}_{-0.06}\times10^{-2} $  & $ 3.8^{+0.06}_{-0.06}\times10^{-2} $ \\
{\footnotesize $\geq 1$ track with no missing middle hits                                     } &  $ 5.4^{+0.10}_{-0.10}\times10^{-2} $  & $ 4.8^{+0.06}_{-0.06}\times10^{-2} $  &  $ 4.7^{+0.06}_{-0.06}\times10^{-2} $  & $ 3.8^{+0.06}_{-0.06}\times10^{-2} $ \\
{\footnotesize $\geq 1$ track with relative track isolation $< 5\%$                           } &  $ 4.6^{+0.10}_{-0.10}\times10^{-2} $  & $ 3.8^{+0.06}_{-0.06}\times10^{-2} $  &  $ 3.5^{+0.05}_{-0.05}\times10^{-2} $  & $ 2.9^{+0.05}_{-0.05}\times10^{-2} $ \\
{\footnotesize $\geq 1$ track with $|d_{\text{xy}}| < 0.02\,\text{cm}{}$                      } &  $ 4.6^{+0.10}_{-0.10}\times10^{-2} $  & $ 3.8^{+0.06}_{-0.06}\times10^{-2} $  &  $ 3.5^{+0.05}_{-0.05}\times10^{-2} $  & $ 2.9^{+0.05}_{-0.05}\times10^{-2} $ \\
{\footnotesize $\geq 1$ track with $|d_z| < 0.5\,\text{cm}{}$                                 } &  $ 4.6^{+0.10}_{-0.10}\times10^{-2} $  & $ 3.8^{+0.06}_{-0.06}\times10^{-2} $  &  $ 3.5^{+0.05}_{-0.05}\times10^{-2} $  & $ 2.9^{+0.05}_{-0.05}\times10^{-2} $ \\
{\footnotesize $\geq 1$ track with $\Delta R (\text{track}, \text{jet}) > 0.5$                } &  $ 4.5^{+0.10}_{-0.10}\times10^{-2} $  & $ 3.7^{+0.06}_{-0.06}\times10^{-2} $  &  $ 3.4^{+0.05}_{-0.05}\times10^{-2} $  & $ 2.8^{+0.05}_{-0.05}\times10^{-2} $ \\
{\footnotesize $\geq 1$ track with $\Delta R (\text{track}, \text{electron}) > 0.15$          } &  $ 4.0^{+0.09}_{-0.09}\times10^{-2} $  & $ 3.7^{+0.06}_{-0.06}\times10^{-2} $  &  $ 3.4^{+0.05}_{-0.05}\times10^{-2} $  & $ 2.8^{+0.05}_{-0.05}\times10^{-2} $ \\
{\footnotesize $\geq 1$ track with $\Delta R (\text{track}, \text{muon}) > 0.15$              } &  $ 1.7^{+0.06}_{-0.06}\times10^{-2} $  & $ 2.5^{+0.05}_{-0.05}\times10^{-2} $  &  $ 2.3^{+0.04}_{-0.04}\times10^{-2} $  & $ 1.9^{+0.04}_{-0.04}\times10^{-2} $ \\
{\footnotesize $\geq 1$ track with $\Delta R (\text{track}, \tau_{\text{h}}) > 0.15$          } &  $ 1.7^{+0.06}_{-0.06}\times10^{-2} $  & $ 2.5^{+0.05}_{-0.05}\times10^{-2} $  &  $ 2.3^{+0.04}_{-0.04}\times10^{-2} $  & $ 1.9^{+0.04}_{-0.04}\times10^{-2} $ \\
{\footnotesize $\geq 1$ track with $E_{\text{calo}} < 10\,\text{GeV}{}$                       } &  $ 1.6^{+0.06}_{-0.06}\times10^{-2} $  & $ 2.5^{+0.05}_{-0.05}\times10^{-2} $  &  $ 2.3^{+0.04}_{-0.04}\times10^{-2} $  & $ 1.9^{+0.04}_{-0.04}\times10^{-2} $ \\
{\footnotesize $\geq 1$ track with $\geq 3$ missing outer hits                                } &  $ 5.4^{+0.33}_{-0.33}\times10^{-3} $  & $ 5.7^{+0.22}_{-0.22}\times10^{-3} $  &  $ 5.1^{+0.21}_{-0.21}\times10^{-3} $  & $ 4.4^{+0.19}_{-0.19}\times10^{-3} $ \\
{\footnotesize $\geq 1$ track with $4$ layers                                                 } &  $ 8.1^{+1.38}_{-1.38}\times10^{-4} $  & $ 6.9^{+0.77}_{-0.77}\times10^{-4} $  &  $ 6.8^{+0.77}_{-0.77}\times10^{-4} $  & $ 5.9^{+0.71}_{-0.71}\times10^{-4} $ \\
{\footnotesize $\geq 1$ track with $5$ layers                                                 } &  $ 6.6^{+1.21}_{-1.21}\times10^{-4} $  & $ 8.4^{+0.85}_{-0.85}\times10^{-4} $  &  $ 6.9^{+0.77}_{-0.77}\times10^{-4} $  & $ 6.9^{+0.77}_{-0.77}\times10^{-4} $ \\
{\footnotesize $\geq 1$ track with $\geq 6$ layers                                            } &  $ 4.0^{+0.29}_{-0.29}\times10^{-3} $  & $ 4.1^{+0.19}_{-0.19}\times10^{-3} $  &  $ 3.6^{+0.18}_{-0.18}\times10^{-3} $  & $ 3.0^{+0.16}_{-0.16}\times10^{-3} $ \\
\hline\hline\end{tabular}
\caption{\label{TAB:MERGING}Comparison of jet matching/merging approaches in cutflows for 700 GeV, 1000 cm winos, region 2017}
\end{table}

%\subsection{Comparison of \MG matching with data}

%In this appendix we list a comparison between \HA and the provided \HEPData information for both $300$ GeV and $700$ GeV winos for the signal region 2018A. A complete set of cutflow comparisons are available online. 

\begin{table}\centering
\begin{tabular}{|c||c|c||c|c||} \hline\hline
 & \multicolumn{2}{|c||}{300 GeV, 10cm} & \multicolumn{2}{|c||}{700 GeV, 10cm} \\ Cut & $\epsilon_i^{\rm CMS}$ & $\epsilon_i^{\rm sim}$, HEPMC &$\epsilon_i^{\rm CMS}$ & $\epsilon_i^{\rm sim}$, HEPMC  \\\hline \hline
{\footnotesize total   } &  $ 1.0^{+0.00}_{-0.00} $  & $ 1.0^{+0.00}_{-0.00} $  &  $ 1.0^{+0.00}_{-0.00} $  & $ 1.0^{+0.00}_{-0.00} $ \\
{\footnotesize trigger } &  $ 9.1^{+0.13}_{-0.13}\times10^{-2} $  & $ 9.2^{+0.09}_{-0.09}\times10^{-2} $  &  $ 1.5^{+0.02}_{-0.02}\times10^{-1} $  & $ 1.5^{+0.01}_{-0.01}\times10^{-1} $ \\
{\footnotesize passes $p_{\text{T}}^{\text{miss}}$ filters } &  $ 9.1^{+0.13}_{-0.13}\times10^{-2} $  & $ 9.2^{+0.09}_{-0.09}\times10^{-2} $  &  $ 1.4^{+0.02}_{-0.02}\times10^{-1} $  & $ 1.5^{+0.01}_{-0.01}\times10^{-1} $ \\
{\footnotesize $p_{\text{T}}^{\text{miss}} > 120\,\text{GeV}{}$ } &  $ 8.9^{+0.13}_{-0.13}\times10^{-2} $  & $ 9.2^{+0.09}_{-0.09}\times10^{-2} $  &  $ 1.4^{+0.02}_{-0.02}\times10^{-1} $  & $ 1.5^{+0.01}_{-0.01}\times10^{-1} $ \\
{\footnotesize $\geq 1$ jet with $p_{\text{T}} > 110\,\text{GeV}{}$ and $|\eta| < 2.4$ } &  $ 8.0^{+0.13}_{-0.13}\times10^{-2} $  & $ 7.5^{+0.09}_{-0.09}\times10^{-2} $  &  $ 1.3^{+0.02}_{-0.02}\times10^{-1} $  & $ 1.3^{+0.01}_{-0.01}\times10^{-1} $ \\
{\footnotesize ==0 pairs of jets with $\Delta\phi_{\text{jet, jet}} > 2.5$ } &  $ 7.0^{+0.12}_{-0.12}\times10^{-2} $  & $ 6.3^{+0.08}_{-0.08}\times10^{-2} $  &  $ 1.1^{+0.01}_{-0.01}\times10^{-1} $  & $ 1.1^{+0.01}_{-0.01}\times10^{-1} $ \\
{\footnotesize $|\Delta\phi(\mbox{leading jet}, \vec{p}_{\text{T}}^{\text{miss}})| > 0.5$ } &  $ 7.0^{+0.12}_{-0.12}\times10^{-2} $  & $ 6.3^{+0.08}_{-0.08}\times10^{-2} $  &  $ 1.1^{+0.01}_{-0.01}\times10^{-1} $  & $ 1.1^{+0.01}_{-0.01}\times10^{-1} $ \\
{\footnotesize $\geq 1$ track with $|\eta| < 2.1$ } &  $ 6.8^{+0.12}_{-0.12}\times10^{-2} $  & $ 6.3^{+0.08}_{-0.08}\times10^{-2} $  &  $ 1.1^{+0.01}_{-0.01}\times10^{-1} $  & $ 1.1^{+0.01}_{-0.01}\times10^{-1} $ \\
{\footnotesize $\geq 1$ track with $p_{\text{T}} > 55\,\text{GeV}{}$ } &  $ 3.2^{+0.08}_{-0.08}\times10^{-2} $  & $ 3.0^{+0.06}_{-0.06}\times10^{-2} $  &  $ 4.7^{+0.10}_{-0.10}\times10^{-2} $  & $ 4.7^{+0.06}_{-0.06}\times10^{-2} $ \\
{\footnotesize $\geq 1$ track passing fiducial selections } &  $ 2.0^{+0.06}_{-0.06}\times10^{-2} $  & $ 2.3^{+0.05}_{-0.05}\times10^{-2} $  &  $ 3.1^{+0.08}_{-0.08}\times10^{-2} $  & $ 3.6^{+0.05}_{-0.05}\times10^{-2} $ \\
{\footnotesize $\geq 1$ track with $\geq 4$ pixel hits } &  $ 1.1^{+0.05}_{-0.05}\times10^{-2} $  & $ 1.7^{+0.04}_{-0.04}\times10^{-2} $  &  $ 1.7^{+0.06}_{-0.06}\times10^{-2} $  & $ 2.6^{+0.05}_{-0.05}\times10^{-2} $ \\
{\footnotesize $\geq 1$ track with no missing inner hits } &  $ 1.1^{+0.05}_{-0.05}\times10^{-2} $  & $ 1.3^{+0.04}_{-0.04}\times10^{-2} $  &  $ 1.7^{+0.06}_{-0.06}\times10^{-2} $  & $ 2.0^{+0.04}_{-0.04}\times10^{-2} $ \\
{\footnotesize $\geq 1$ track with no missing middle hits } &  $ 1.0^{+0.05}_{-0.05}\times10^{-2} $  & $ 1.3^{+0.04}_{-0.04}\times10^{-2} $  &  $ 1.5^{+0.05}_{-0.05}\times10^{-2} $  & $ 2.0^{+0.04}_{-0.04}\times10^{-2} $ \\
{\footnotesize $\geq 1$ track with relative track isolation $< 5\%$ } &  $ 5.1^{+0.32}_{-0.32}\times10^{-3} $  & $ 6.2^{+0.26}_{-0.26}\times10^{-3} $  &  $ 5.3^{+0.34}_{-0.34}\times10^{-3} $  & $ 6.2^{+0.23}_{-0.23}\times10^{-3} $ \\
{\footnotesize $\geq 1$ track with $|d_{\text{xy}}| < 0.02\,\text{cm}{}$ } &  $ 5.1^{+0.32}_{-0.32}\times10^{-3} $  & $ 6.2^{+0.26}_{-0.26}\times10^{-3} $  &  $ 5.1^{+0.34}_{-0.34}\times10^{-3} $  & $ 6.2^{+0.23}_{-0.23}\times10^{-3} $ \\
{\footnotesize $\geq 1$ track with $|d_z| < 0.5\,\text{cm}{}$ } &  $ 5.1^{+0.32}_{-0.32}\times10^{-3} $  & $ 6.2^{+0.26}_{-0.26}\times10^{-3} $  &  $ 5.1^{+0.31}_{-0.31}\times10^{-3} $  & $ 6.2^{+0.23}_{-0.23}\times10^{-3} $ \\
{\footnotesize $\geq 1$ track with $\Delta R (\text{track}, \text{jet}) > 0.5$ } &  $ 5.0^{+0.32}_{-0.32}\times10^{-3} $  & $ 6.1^{+0.25}_{-0.25}\times10^{-3} $  &  $ 4.9^{+0.31}_{-0.31}\times10^{-3} $  & $ 6.1^{+0.23}_{-0.23}\times10^{-3} $ \\
{\footnotesize $\geq 1$ track with $\Delta R (\text{track}, \text{electron}) > 0.15$ } &  $ 4.9^{+0.31}_{-0.31}\times10^{-3} $  & $ 6.1^{+0.25}_{-0.25}\times10^{-3} $  &  $ 4.7^{+0.31}_{-0.31}\times10^{-3} $  & $ 6.1^{+0.23}_{-0.23}\times10^{-3} $ \\
{\footnotesize $\geq 1$ track with $\Delta R (\text{track}, \text{muon}) > 0.15$ } &  $ 4.9^{+0.31}_{-0.31}\times10^{-3} $  & $ 6.1^{+0.25}_{-0.25}\times10^{-3} $  &  $ 4.7^{+0.31}_{-0.31}\times10^{-3} $  & $ 6.1^{+0.23}_{-0.23}\times10^{-3} $ \\
{\footnotesize $\geq 1$ track with $\Delta R (\text{track}, \tau_{\text{h}}) > 0.15$ } &  $ 4.9^{+0.31}_{-0.31}\times10^{-3} $  & $ 6.1^{+0.25}_{-0.25}\times10^{-3} $  &  $ 4.7^{+0.31}_{-0.31}\times10^{-3} $  & $ 6.1^{+0.23}_{-0.23}\times10^{-3} $ \\
{\footnotesize $\geq 1$ track with $E_{\text{calo}} < 10\,\text{GeV}{}$ } &  $ 4.8^{+0.31}_{-0.31}\times10^{-3} $  & $ 6.1^{+0.25}_{-0.25}\times10^{-3} $  &  $ 4.7^{+0.31}_{-0.31}\times10^{-3} $  & $ 6.1^{+0.23}_{-0.23}\times10^{-3} $ \\
{\footnotesize $\geq 1$ track with $\geq 3$ missing outer hits } &  $ 4.8^{+0.31}_{-0.31}\times10^{-3} $  & $ 5.9^{+0.25}_{-0.25}\times10^{-3} $  &  $ 4.6^{+0.31}_{-0.31}\times10^{-3} $  & $ 6.1^{+0.23}_{-0.23}\times10^{-3} $ \\
{\footnotesize $\geq 1$ track $4$ layers } &  $ 2.6^{+0.23}_{-0.23}\times10^{-3} $  & $ 2.5^{+0.16}_{-0.16}\times10^{-3} $  &  $ 2.8^{+0.24}_{-0.24}\times10^{-3} $  & $ 3.4^{+0.17}_{-0.17}\times10^{-3} $ \\
{\footnotesize $\geq 1$ track $5$ layers } &  $ 1.1^{+0.15}_{-0.15}\times10^{-3} $  & $ 1.3^{+0.12}_{-0.12}\times10^{-3} $  &  $ 9.2^{+1.36}_{-1.36}\times10^{-4} $  & $ 1.3^{+0.11}_{-0.11}\times10^{-3} $ \\
{\footnotesize $\geq 1$ track with $\geq 6$ layers } &  $ 1.1^{+0.15}_{-0.15}\times10^{-3} $  & $ 1.9^{+0.14}_{-0.14}\times10^{-3} $  &  $ 9.5^{+1.36}_{-1.36}\times10^{-4} $  & $ 1.2^{+0.10}_{-0.10}\times10^{-3} $ \\
\hline\hline\end{tabular}
\caption{Cutflow comparison for signal region 2018A and decay length 10cm, for $300$ GeV and $700$ GeV winos.}\label{TAB:10cm}
\end{table}
\begin{table}\centering
\begin{tabular}{|c||c|c||c|c||} \hline\hline
 & \multicolumn{2}{|c||}{300 GeV, 100cm} & \multicolumn{2}{|c||}{700 GeV, 100cm} \\ Cut & $\epsilon_i^{\rm CMS}$ & $\epsilon_i^{\rm sim}$, HEPMC &$\epsilon_i^{\rm CMS}$ & $\epsilon_i^{\rm sim}$, HEPMC  \\\hline \hline
{\footnotesize total   } &  $ 1.0^{+0.00}_{-0.00} $  & $ 1.0^{+0.00}_{-0.00} $  &  $ 1.0^{+0.01}_{-0.01} $  & $ 1.0^{+0.00}_{-0.00} $ \\
{\footnotesize trigger } &  $ 9.5^{+0.14}_{-0.14}\times10^{-2} $  & $ 9.3^{+0.09}_{-0.09}\times10^{-2} $  &  $ 1.5^{+0.02}_{-0.02}\times10^{-1} $  & $ 1.5^{+0.01}_{-0.01}\times10^{-1} $ \\
{\footnotesize passes $p_{\text{T}}^{\text{miss}}$ filters } &  $ 9.5^{+0.14}_{-0.14}\times10^{-2} $  & $ 9.3^{+0.09}_{-0.09}\times10^{-2} $  &  $ 1.5^{+0.02}_{-0.02}\times10^{-1} $  & $ 1.5^{+0.01}_{-0.01}\times10^{-1} $ \\
{\footnotesize $p_{\text{T}}^{\text{miss}} > 120\,\text{GeV}{}$ } &  $ 9.3^{+0.14}_{-0.14}\times10^{-2} $  & $ 9.3^{+0.09}_{-0.09}\times10^{-2} $  &  $ 1.5^{+0.02}_{-0.02}\times10^{-1} $  & $ 1.5^{+0.01}_{-0.01}\times10^{-1} $ \\
{\footnotesize $\geq 1$ jet with $p_{\text{T}} > 110\,\text{GeV}{}$ and $|\eta| < 2.4$ } &  $ 8.2^{+0.13}_{-0.13}\times10^{-2} $  & $ 7.5^{+0.09}_{-0.09}\times10^{-2} $  &  $ 1.3^{+0.02}_{-0.02}\times10^{-1} $  & $ 1.3^{+0.01}_{-0.01}\times10^{-1} $ \\
{\footnotesize ==0 pairs of jets with $\Delta\phi_{\text{jet, jet}} > 2.5$ } &  $ 7.1^{+0.12}_{-0.12}\times10^{-2} $  & $ 6.3^{+0.08}_{-0.08}\times10^{-2} $  &  $ 1.1^{+0.02}_{-0.02}\times10^{-1} $  & $ 1.1^{+0.01}_{-0.01}\times10^{-1} $ \\
{\footnotesize $|\Delta\phi(\mbox{leading jet}, \vec{p}_{\text{T}}^{\text{miss}})| > 0.5$ } &  $ 7.1^{+0.12}_{-0.12}\times10^{-2} $  & $ 6.3^{+0.08}_{-0.08}\times10^{-2} $  &  $ 1.1^{+0.02}_{-0.02}\times10^{-1} $  & $ 1.1^{+0.01}_{-0.01}\times10^{-1} $ \\
{\footnotesize $\geq 1$ track with $|\eta| < 2.1$ } &  $ 7.0^{+0.12}_{-0.12}\times10^{-2} $  & $ 6.3^{+0.08}_{-0.08}\times10^{-2} $  &  $ 1.1^{+0.02}_{-0.02}\times10^{-1} $  & $ 1.1^{+0.01}_{-0.01}\times10^{-1} $ \\
{\footnotesize $\geq 1$ track with $p_{\text{T}} > 55\,\text{GeV}{}$ } &  $ 5.5^{+0.11}_{-0.11}\times10^{-2} $  & $ 5.2^{+0.07}_{-0.07}\times10^{-2} $  &  $ 8.7^{+0.17}_{-0.17}\times10^{-2} $  & $ 9.1^{+0.08}_{-0.08}\times10^{-2} $ \\
{\footnotesize $\geq 1$ track passing fiducial selections } &  $ 3.8^{+0.09}_{-0.09}\times10^{-2} $  & $ 4.2^{+0.07}_{-0.07}\times10^{-2} $  &  $ 6.1^{+0.17}_{-0.17}\times10^{-2} $  & $ 7.5^{+0.08}_{-0.08}\times10^{-2} $ \\
{\footnotesize $\geq 1$ track with $\geq 4$ pixel hits } &  $ 2.5^{+0.07}_{-0.07}\times10^{-2} $  & $ 3.2^{+0.06}_{-0.06}\times10^{-2} $  &  $ 3.9^{+0.14}_{-0.14}\times10^{-2} $  & $ 5.7^{+0.07}_{-0.07}\times10^{-2} $ \\
{\footnotesize $\geq 1$ track with no missing inner hits } &  $ 2.5^{+0.07}_{-0.07}\times10^{-2} $  & $ 2.4^{+0.05}_{-0.05}\times10^{-2} $  &  $ 3.9^{+0.14}_{-0.14}\times10^{-2} $  & $ 4.5^{+0.06}_{-0.06}\times10^{-2} $ \\
{\footnotesize $\geq 1$ track with no missing middle hits } &  $ 2.2^{+0.07}_{-0.07}\times10^{-2} $  & $ 2.4^{+0.05}_{-0.05}\times10^{-2} $  &  $ 3.6^{+0.14}_{-0.14}\times10^{-2} $  & $ 4.5^{+0.06}_{-0.06}\times10^{-2} $ \\
{\footnotesize $\geq 1$ track with relative track isolation $< 5\%$ } &  $ 1.8^{+0.06}_{-0.06}\times10^{-2} $  & $ 1.7^{+0.04}_{-0.04}\times10^{-2} $  &  $ 2.9^{+0.11}_{-0.11}\times10^{-2} $  & $ 3.4^{+0.05}_{-0.05}\times10^{-2} $ \\
{\footnotesize $\geq 1$ track with $|d_{\text{xy}}| < 0.02\,\text{cm}{}$ } &  $ 1.8^{+0.06}_{-0.06}\times10^{-2} $  & $ 1.7^{+0.04}_{-0.04}\times10^{-2} $  &  $ 2.9^{+0.11}_{-0.11}\times10^{-2} $  & $ 3.4^{+0.05}_{-0.05}\times10^{-2} $ \\
{\footnotesize $\geq 1$ track with $|d_z| < 0.5\,\text{cm}{}$ } &  $ 1.8^{+0.06}_{-0.06}\times10^{-2} $  & $ 1.7^{+0.04}_{-0.04}\times10^{-2} $  &  $ 2.9^{+0.11}_{-0.11}\times10^{-2} $  & $ 3.4^{+0.05}_{-0.05}\times10^{-2} $ \\
{\footnotesize $\geq 1$ track with $\Delta R (\text{track}, \text{jet}) > 0.5$ } &  $ 1.8^{+0.06}_{-0.06}\times10^{-2} $  & $ 1.7^{+0.04}_{-0.04}\times10^{-2} $  &  $ 2.8^{+0.11}_{-0.11}\times10^{-2} $  & $ 3.3^{+0.05}_{-0.05}\times10^{-2} $ \\
{\footnotesize $\geq 1$ track with $\Delta R (\text{track}, \text{electron}) > 0.15$ } &  $ 1.7^{+0.06}_{-0.06}\times10^{-2} $  & $ 1.7^{+0.04}_{-0.04}\times10^{-2} $  &  $ 2.6^{+0.10}_{-0.10}\times10^{-2} $  & $ 3.3^{+0.05}_{-0.05}\times10^{-2} $ \\
{\footnotesize $\geq 1$ track with $\Delta R (\text{track}, \text{muon}) > 0.15$ } &  $ 1.6^{+0.06}_{-0.06}\times10^{-2} $  & $ 1.7^{+0.04}_{-0.04}\times10^{-2} $  &  $ 2.6^{+0.10}_{-0.10}\times10^{-2} $  & $ 3.3^{+0.05}_{-0.05}\times10^{-2} $ \\
{\footnotesize $\geq 1$ track with $\Delta R (\text{track}, \tau_{\text{h}}) > 0.15$ } &  $ 1.6^{+0.06}_{-0.06}\times10^{-2} $  & $ 1.7^{+0.04}_{-0.04}\times10^{-2} $  &  $ 2.6^{+0.10}_{-0.10}\times10^{-2} $  & $ 3.3^{+0.05}_{-0.05}\times10^{-2} $ \\
{\footnotesize $\geq 1$ track with $E_{\text{calo}} < 10\,\text{GeV}{}$ } &  $ 1.6^{+0.06}_{-0.06}\times10^{-2} $  & $ 1.7^{+0.04}_{-0.04}\times10^{-2} $  &  $ 2.5^{+0.10}_{-0.10}\times10^{-2} $  & $ 3.3^{+0.05}_{-0.05}\times10^{-2} $ \\
{\footnotesize $\geq 1$ track with $\geq 3$ missing outer hits } &  $ 8.8^{+0.42}_{-0.42}\times10^{-3} $  & $ 9.6^{+0.32}_{-0.32}\times10^{-3} $  &  $ 1.7^{+0.08}_{-0.08}\times10^{-2} $  & $ 2.2^{+0.04}_{-0.04}\times10^{-2} $ \\
{\footnotesize $\geq 1$ track $4$ layers } &  $ 1.7^{+0.19}_{-0.19}\times10^{-3} $  & $ 1.7^{+0.13}_{-0.13}\times10^{-3} $  &  $ 3.8^{+0.37}_{-0.37}\times10^{-3} $  & $ 4.2^{+0.19}_{-0.19}\times10^{-3} $ \\
{\footnotesize $\geq 1$ track $5$ layers } &  $ 1.6^{+0.18}_{-0.18}\times10^{-3} $  & $ 1.3^{+0.12}_{-0.12}\times10^{-3} $  &  $ 2.8^{+0.34}_{-0.34}\times10^{-3} $  & $ 3.6^{+0.18}_{-0.18}\times10^{-3} $ \\
{\footnotesize $\geq 1$ track with $\geq 6$ layers } &  $ 5.7^{+0.34}_{-0.34}\times10^{-3} $  & $ 6.5^{+0.26}_{-0.26}\times10^{-3} $  &  $ 1.1^{+0.06}_{-0.06}\times10^{-2} $  & $ 1.4^{+0.03}_{-0.03}\times10^{-2} $ \\
\hline\hline\end{tabular}
\caption{Cutflow comparison for signal region 2018A and decay length 100cm, for $300$ GeV and $700$ GeV winos.}\label{TAB:100cm}
\end{table}
\begin{table}\centering
\begin{tabular}{|c||c|c||c|c||} \hline\hline
 & \multicolumn{2}{|c||}{300 GeV, 1000cm} & \multicolumn{2}{|c||}{700 GeV, 1000cm} \\ Cut & $\epsilon_i^{\rm CMS}$ & $\epsilon_i^{\rm sim}$, HEPMC &$\epsilon_i^{\rm CMS}$ & $\epsilon_i^{\rm sim}$, HEPMC  \\\hline \hline
{\footnotesize total   } &  $ 1.0^{+0.00}_{-0.00} $  & $ 1.0^{+0.00}_{-0.00} $  &  $ 1.0^{+0.00}_{-0.00} $  & $ 1.0^{+0.00}_{-0.00} $ \\
{\footnotesize trigger } &  $ 9.8^{+0.14}_{-0.14}\times10^{-2} $  & $ 1.1^{+0.01}_{-0.01}\times10^{-1} $  &  $ 1.6^{+0.02}_{-0.02}\times10^{-1} $  & $ 1.8^{+0.01}_{-0.01}\times10^{-1} $ \\
{\footnotesize passes $p_{\text{T}}^{\text{miss}}$ filters } &  $ 9.8^{+0.14}_{-0.14}\times10^{-2} $  & $ 1.1^{+0.01}_{-0.01}\times10^{-1} $  &  $ 1.6^{+0.02}_{-0.02}\times10^{-1} $  & $ 1.8^{+0.01}_{-0.01}\times10^{-1} $ \\
{\footnotesize $p_{\text{T}}^{\text{miss}} > 120\,\text{GeV}{}$ } &  $ 9.5^{+0.14}_{-0.14}\times10^{-2} $  & $ 1.1^{+0.01}_{-0.01}\times10^{-1} $  &  $ 1.5^{+0.02}_{-0.02}\times10^{-1} $  & $ 1.8^{+0.01}_{-0.01}\times10^{-1} $ \\
{\footnotesize $\geq 1$ jet with $p_{\text{T}} > 110\,\text{GeV}{}$ and $|\eta| < 2.4$ } &  $ 8.2^{+0.13}_{-0.13}\times10^{-2} $  & $ 7.5^{+0.09}_{-0.09}\times10^{-2} $  &  $ 1.4^{+0.02}_{-0.02}\times10^{-1} $  & $ 1.4^{+0.01}_{-0.01}\times10^{-1} $ \\
{\footnotesize ==0 pairs of jets with $\Delta\phi_{\text{jet, jet}} > 2.5$ } &  $ 7.1^{+0.12}_{-0.12}\times10^{-2} $  & $ 6.3^{+0.08}_{-0.08}\times10^{-2} $  &  $ 1.2^{+0.01}_{-0.01}\times10^{-1} $  & $ 1.2^{+0.01}_{-0.01}\times10^{-1} $ \\
{\footnotesize $|\Delta\phi(\mbox{leading jet}, \vec{p}_{\text{T}}^{\text{miss}})| > 0.5$ } &  $ 7.1^{+0.12}_{-0.12}\times10^{-2} $  & $ 6.0^{+0.08}_{-0.08}\times10^{-2} $  &  $ 1.2^{+0.01}_{-0.01}\times10^{-1} $  & $ 1.2^{+0.01}_{-0.01}\times10^{-1} $ \\
{\footnotesize $\geq 1$ track with $|\eta| < 2.1$ } &  $ 7.0^{+0.12}_{-0.12}\times10^{-2} $  & $ 6.0^{+0.08}_{-0.08}\times10^{-2} $  &  $ 1.2^{+0.01}_{-0.01}\times10^{-1} $  & $ 1.2^{+0.01}_{-0.01}\times10^{-1} $ \\
{\footnotesize $\geq 1$ track with $p_{\text{T}} > 55\,\text{GeV}{}$ } &  $ 6.1^{+0.11}_{-0.11}\times10^{-2} $  & $ 5.4^{+0.07}_{-0.07}\times10^{-2} $  &  $ 1.0^{+0.01}_{-0.01}\times10^{-1} $  & $ 1.1^{+0.01}_{-0.01}\times10^{-1} $ \\
{\footnotesize $\geq 1$ track passing fiducial selections } &  $ 4.2^{+0.09}_{-0.09}\times10^{-2} $  & $ 4.4^{+0.07}_{-0.07}\times10^{-2} $  &  $ 7.5^{+0.14}_{-0.14}\times10^{-2} $  & $ 8.7^{+0.08}_{-0.08}\times10^{-2} $ \\
{\footnotesize $\geq 1$ track with $\geq 4$ pixel hits } &  $ 2.9^{+0.08}_{-0.08}\times10^{-2} $  & $ 3.4^{+0.06}_{-0.06}\times10^{-2} $  &  $ 5.3^{+0.10}_{-0.10}\times10^{-2} $  & $ 7.0^{+0.07}_{-0.07}\times10^{-2} $ \\
{\footnotesize $\geq 1$ track with no missing inner hits } &  $ 2.9^{+0.08}_{-0.08}\times10^{-2} $  & $ 2.4^{+0.05}_{-0.05}\times10^{-2} $  &  $ 5.2^{+0.10}_{-0.10}\times10^{-2} $  & $ 4.8^{+0.06}_{-0.06}\times10^{-2} $ \\
{\footnotesize $\geq 1$ track with no missing middle hits } &  $ 2.5^{+0.07}_{-0.07}\times10^{-2} $  & $ 2.4^{+0.05}_{-0.05}\times10^{-2} $  &  $ 4.6^{+0.10}_{-0.10}\times10^{-2} $  & $ 4.8^{+0.06}_{-0.06}\times10^{-2} $ \\
{\footnotesize $\geq 1$ track with relative track isolation $< 5\%$ } &  $ 2.1^{+0.06}_{-0.06}\times10^{-2} $  & $ 1.8^{+0.04}_{-0.04}\times10^{-2} $  &  $ 3.8^{+0.10}_{-0.10}\times10^{-2} $  & $ 3.8^{+0.06}_{-0.06}\times10^{-2} $ \\
{\footnotesize $\geq 1$ track with $|d_{\text{xy}}| < 0.02\,\text{cm}{}$ } &  $ 2.1^{+0.06}_{-0.06}\times10^{-2} $  & $ 1.8^{+0.04}_{-0.04}\times10^{-2} $  &  $ 3.8^{+0.10}_{-0.10}\times10^{-2} $  & $ 3.8^{+0.06}_{-0.06}\times10^{-2} $ \\
{\footnotesize $\geq 1$ track with $|d_z| < 0.5\,\text{cm}{}$ } &  $ 2.1^{+0.06}_{-0.06}\times10^{-2} $  & $ 1.8^{+0.04}_{-0.04}\times10^{-2} $  &  $ 3.8^{+0.10}_{-0.10}\times10^{-2} $  & $ 3.8^{+0.06}_{-0.06}\times10^{-2} $ \\
{\footnotesize $\geq 1$ track with $\Delta R (\text{track}, \text{jet}) > 0.5$ } &  $ 2.0^{+0.06}_{-0.06}\times10^{-2} $  & $ 1.7^{+0.04}_{-0.04}\times10^{-2} $  &  $ 3.8^{+0.10}_{-0.10}\times10^{-2} $  & $ 3.7^{+0.06}_{-0.06}\times10^{-2} $ \\
{\footnotesize $\geq 1$ track with $\Delta R (\text{track}, \text{electron}) > 0.15$ } &  $ 1.9^{+0.06}_{-0.06}\times10^{-2} $  & $ 1.7^{+0.04}_{-0.04}\times10^{-2} $  &  $ 3.3^{+0.08}_{-0.08}\times10^{-2} $  & $ 3.7^{+0.06}_{-0.06}\times10^{-2} $ \\
{\footnotesize $\geq 1$ track with $\Delta R (\text{track}, \text{muon}) > 0.15$ } &  $ 6.6^{+0.37}_{-0.37}\times10^{-3} $  & $ 9.5^{+0.32}_{-0.32}\times10^{-3} $  &  $ 1.4^{+0.05}_{-0.05}\times10^{-2} $  & $ 2.5^{+0.05}_{-0.05}\times10^{-2} $ \\
{\footnotesize $\geq 1$ track with $\Delta R (\text{track}, \tau_{\text{h}}) > 0.15$ } &  $ 6.6^{+0.37}_{-0.37}\times10^{-3} $  & $ 9.5^{+0.32}_{-0.32}\times10^{-3} $  &  $ 1.4^{+0.05}_{-0.05}\times10^{-2} $  & $ 2.5^{+0.05}_{-0.05}\times10^{-2} $ \\
{\footnotesize $\geq 1$ track with $E_{\text{calo}} < 10\,\text{GeV}{}$ } &  $ 6.5^{+0.36}_{-0.36}\times10^{-3} $  & $ 9.5^{+0.32}_{-0.32}\times10^{-3} $  &  $ 1.3^{+0.05}_{-0.05}\times10^{-2} $  & $ 2.5^{+0.05}_{-0.05}\times10^{-2} $ \\
{\footnotesize $\geq 1$ track with $\geq 3$ missing outer hits } &  $ 1.9^{+0.20}_{-0.20}\times10^{-3} $  & $ 1.9^{+0.14}_{-0.14}\times10^{-3} $  &  $ 4.6^{+0.31}_{-0.31}\times10^{-3} $  & $ 5.7^{+0.22}_{-0.22}\times10^{-3} $ \\
{\footnotesize $\geq 1$ track $4$ layers } &  $ 3.7^{+0.86}_{-0.86}\times10^{-4} $  & $ 2.1^{+0.48}_{-0.48}\times10^{-4} $  &  $ 7.1^{+1.36}_{-1.36}\times10^{-4} $  & $ 6.7^{+0.76}_{-0.76}\times10^{-4} $ \\
{\footnotesize $\geq 1$ track $5$ layers } &  $ 2.4^{+0.71}_{-0.71}\times10^{-4} $  & $ 1.6^{+0.42}_{-0.42}\times10^{-4} $  &  $ 4.8^{+1.02}_{-1.02}\times10^{-4} $  & $ 8.7^{+0.87}_{-0.87}\times10^{-4} $ \\
{\footnotesize $\geq 1$ track with $\geq 6$ layers } &  $ 1.3^{+0.16}_{-0.16}\times10^{-3} $  & $ 1.5^{+0.13}_{-0.13}\times10^{-3} $  &  $ 3.4^{+0.27}_{-0.27}\times10^{-3} $  & $ 4.0^{+0.18}_{-0.18}\times10^{-3} $ \\
\hline\hline\end{tabular}
\caption{Cutflow comparison for signal region 2018A and decay length 1000cm, for $300$ GeV and $700$ GeV winos.}\label{TAB:1000cm}
\end{table}

\newpage
 \bibliographystyle{h-physrev}
\bibliography{literature}

\end{document}